\documentclass[twocolumn,aps,prb,superscriptaddress,nofootinbib,floatfix]{revtex4-2}

\usepackage[T1]{fontenc}
\usepackage{amsmath,amssymb,amsfonts}
\usepackage{graphicx}
\usepackage{xcolor}
\usepackage{booktabs}
\usepackage{textcomp}
\usepackage{caption}
\usepackage{subcaption}
\usepackage{enumerate}
\usepackage{soul}
\usepackage{verbatim}
\usepackage{ragged2e}

\renewcommand{\thesection}{\arabic{section}}
\renewcommand{\thesubsection}{\arabic{section}.\arabic{subsection}}
\renewcommand{\thesubsubsection}{\arabic{section}.\arabic{subsection}.\arabic{subsubsection}}

\newenvironment{widetab}{\begin{table}}{\end{table}}
\newcommand{\beginwide}{\begin{widetext}}
\newcommand{\myfigwidth}{\columnwidth}

\newcommand{\fillin}[1]{}
\newcommand{\citeneeded}[1]{}
\newcommand{\sz}[1]{}
\newcommand{\jm}[1]{}
\newcommand{\jy}[1]{}
\newcommand{\ar}[1]{}
\providecommand{\backmatter}{}

\usepackage[colorlinks=true,allcolors=blue]{hyperref}

\makeatletter
\renewcommand{\p@subsection}{}
\renewcommand{\p@subsubsection}{}
\makeatother

\begin{document}

\title{Neural-Network Inverse Design of Cavity–Transmon Systems for Bosonic Quantum Computing}
\title{Deep Learning Aided Inverse Design of SRF Cavities and Transmons for Quantum Computation}
\title{Neural-Network Inverse Design of SRF Cavities and Transmons for Bosonic Quantum Computation}

\author{Joseph Yaker}
\affiliation{Superconducting and Quantum Materials System Center (SQMS), Fermi National Accelerator Laboratory, Batavia, 60510, IL, USA}
\affiliation{Applied Physics Program, Northwestern University, Evanston, 60208, IL, USA}

\author{Jovan Markovic}
\affiliation{Superconducting and Quantum Materials System Center (SQMS), Fermi National Accelerator Laboratory, Batavia, 60510, IL, USA}
\affiliation{Department of Physics, University of Cambridge, Cambridge, CB3 0US, UK}

\author{Alessandro Reineri}
\affiliation{Superconducting and Quantum Materials System Center (SQMS), Fermi National Accelerator Laboratory, Batavia, 60510, IL, USA}
\affiliation{Illinois Institute of Technology, 10 W 35th Street, Chicago, 60616, IL, USA}

\author{Doga Murat Kurkcuoglu}
\affiliation{Superconducting and Quantum Materials System Center (SQMS), Fermi National Accelerator Laboratory, Batavia, 60510, IL, USA}

\author{Silvia Zorzetti}
\email{zorzetti@fnal.gov}
\affiliation{Superconducting and Quantum Materials System Center (SQMS), Fermi National Accelerator Laboratory, Batavia, 60510, IL, USA}
\affiliation{Department of Physics and Astronomy, Northwestern University, Evanston, 60208, IL, USA}

\begin{abstract}
Three-dimensional superconducting radio-frequency (SRF) cavities provide exceptionally long-lived electromagnetic modes and, when coupled to nonlinear elements such as transmon qubits, become promising architectures for bosonic quantum information processing. The inverse design of such systems, i.e., recovering device geometries that produce specified electromagnetic and coupling targets, is generally a one-to-many problem. The qubit-cavity coupling strength depends sensitively on both the transmon geometry and its position within the cavity's electromagnetic field. As these systems scale up and their design parameter spaces grow, the cost of conventional iterative simulation becomes prohibitive. We present two deep neural network (DNN) approaches that address this inverse-design problem at complementary levels of the design stack. The first proposes SRF cavity geometries that produce target cavity observables. The second proposes transmon qubit designs that produce target qubit--cavity parameters --- the coupling rate, qubit frequency, and anharmonicity $(g, \nu_q, \alpha)$. The recovered candidate designs match the targets to within $\sim$5\% (cavity) and $\sim$2\% (transmon), confirmed by end-to-end re-simulation. Both approaches map desired device behavior directly to candidate designs, a fast alternative to the iterative simulation studies usually required.

\end{abstract}


\maketitle

\section{Introduction}

Designing electromagnetic devices to meet target performance specifications is a recurring challenge across microwave engineering, photonics, and accelerator physics. The forward problem, i.e., predicting device performance from a given geometry, is routinely tackled with finite-element software packages for electromagnetic and multiphysics simulations. The inverse problem is the recovery of a geometry that produces the desired response. This direction is harder because the mapping to viable solutions is generally one-to-many.  Therefore, with traditional approaches, the cost of repeated simulation grows rapidly with the dimensionality of the design space. Machine learning offers a complementary route and has already been applied to a wide range of problems in the physical sciences \cite{CarleoMLPhysics2019, RadovicHEP2018}, including the rapid solution of complex electromagnetic inverse design problems~\cite{PeurifoyInverse2018, RaissiPINNs2019}. Here, we apply this approach to the design and optimization of superconducting devices for quantum computing and sensing.

Superconducting radio frequency (SRF) cavities are highly efficient microwave resonators, historically developed to accelerate subatomic particles. They have also proven useful in quantum information science, acting as high-$Q$, low-loss resonators used to store and protect quantum states~\cite{PaikCavity2011, romanenko2020three, reagor2016quantum}. Coupled to nonlinear superconducting circuits, i.e., qubits~\cite{KochTransmon2007}, used to control and measure quantum information within, they can achieve very long single-photon lifetimes in a range of \(10\text{--}30\text{ ms}\) ~\cite{KimSQMS2506,ReineriSQMS2308, milul2023superconducting}. These long-lived quantum memories are particularly relevant in the field of high energy physics for quantum simulations of field theories and related applications.

The coupling between SRF cavities and superconducting qubits, most commonly transmons~\cite{KochTransmon2007}, is described by the formalism of three-dimensional circuit quantum electrodynamics (3D cQED)~\cite{PaikCavity2011}. The transmon is dispersively coupled to the cavity environment and, being weakly anharmonic~\cite{KochTransmon2007} (its anharmonicity is small compared to its transition frequency), supplies the Josephson nonlinearity required for quantum control~\cite{PaikCavity2011, KimSQMS2506}. This coupled transmon--cavity system enables the preparation of complex bosonic states~\cite{HeeresSNAP2015}, the realization of bosonic error-correcting codes, and the demonstration of high-fidelity universal gate sets \cite{VlastakisCat2013,HeeresSNAP2015,CampagneGKP2020}. In cavity-based bosonic quantum computing, Fock states of the normal modes of the electromagnetic (EM) field can be utilized as qudits. The EM field is coupled to the transmon through a dipole coupling, and the coupling strength is set by the local electric-field amplitude at the transmon's position. The qudit gates result from pulses applied to the cavity and transmon, and the coupling strength defines the pulse shape. Finding the cavity shape for a given EM field is thus helpful for bosonic quantum computing.

Microwave control drives are applied directly to the transmon, and through the dispersive coupling the cavity modes inherit its nonlinearity, enabling ancilla-mediated control of the bosonic Hilbert space~\cite{HeeresSNAP2015,BlaisDispersive2004}. At the design phase, energy participation ratio (EPR) analysis~\cite{MinevEPR2021, yaker2025quantum} provides a framework for extracting the dispersive Hamiltonian from finite-element eigenmode simulations together with a perturbative expansion of the Josephson nonlinearity. Since the participation ratios are related to the system's geometric features, this method ultimately quantifies how the dispersive Hamiltonian parameters depend on cavity geometry, opening a path towards geometry optimization for specified parameter targets. This dispersive Hamiltonian is governed by a dense set of self- and cross-Kerr interactions whose magnitudes depend nonlinearly on the device geometry, and which must be controlled simultaneously to suppress spurious crosstalk while realizing the target couplings. Capturing this high-dimensional dependence is impractical by manual tuning but is a natural target for data-driven inverse design.

Design optimization workflows have traditionally involved iterative parameterized redesign. Automated optimization methods such as adjoint-based shape determination~\cite{Akcelik2008} and evolutionary search~\cite{KranjcevicRF2019} have been in use for over a decade. A limitation of these approaches is that they tune for only a subset of EM field parameters, such as eigenmodes and electric field distribution at the chip location. As noted, the inverse mapping of the target EM parameters to the adequate geometry is one-to-many, and finding it via parameter sweeps with packages such as COMSOL Multiphysics or Ansys HFSS can be prohibitively costly.  
We present deep neural networks (DNNs) that generate SRF cavity geometries from targeted electromagnetic parameters--specifically, the fundamental mode 
passband width and the axial electric field ($E_z$) profile at the prospective qubit location. 
A second DNN addresses the analogous inverse problem on the transmon qubit side, mapping target qubit--cavity parameters --- the coupling rate, frequency, and anharmonicity $(g, \nu_q, \alpha)$ --- to transmon design variables, i.e., qubit insertion depth, Josephson inductance, and shunt-pad width, $(z_\mathrm{trans}, L_J, w_\mathrm{pad})$. On the transmon side, these target parameters are themselves obtained from finite-element eigenmode simulations processed through the \texttt{pyEPR} implementation of the EPR method~\cite{MinevEPR2021}: the network thus learns to invert an established quantum-device characterization pipeline, coupling a data-driven inverse model directly to the physics-based extraction of the dispersive Hamiltonian. This places the work in the \emph{AI-for-quantum} direction, using machine learning to accelerate the design of superconducting quantum hardware. Its relevance grows as quantum devices scale up and their parameter spaces expand, calling for hardware–software co-design to tailor universal gate sets to specific applications~\cite{you2024crosstalk}.

Section~\ref{sec:background} details cavity QED theory and the dispersive cavity--transmon Hamiltonian. Section~\ref{sec:method} describes geometry parametrization, DNN architectures, and training procedures. Section~\ref{sec:results} reports cavity DNN performance, EPR-extracted coupling parameters for cavity-DNN geometries, and transmon DNN performance. Section~\ref{sec:conclusions} summarizes findings and discusses limitations.

\section{Background}
\label{sec:background}

In cavity QED~\cite{walther2006cavity, BlaisReview2021}, resonant modes of the cavity provide a controllable environment in which quantum states can be prepared, manipulated, and read out with high fidelity through the use of an ancillary transmon. SRF cavities are particularly well suited for this role, with a demonstrated record high photon lifetime, exceeding 2 seconds, enabled by ultra-low inner wall surface resistance that yields an intrinsic quality factor of  $Q_0>10^{10}$~\cite{romanenko2020three}. In this section, we examine common SRF cavity geometries relevant to cavity QED and identify the key figures of merit that characterize their performance.

\subsection{Common Cavity Geometries}
\begin{figure}[t]
\centering
\includegraphics[width=\myfigwidth]{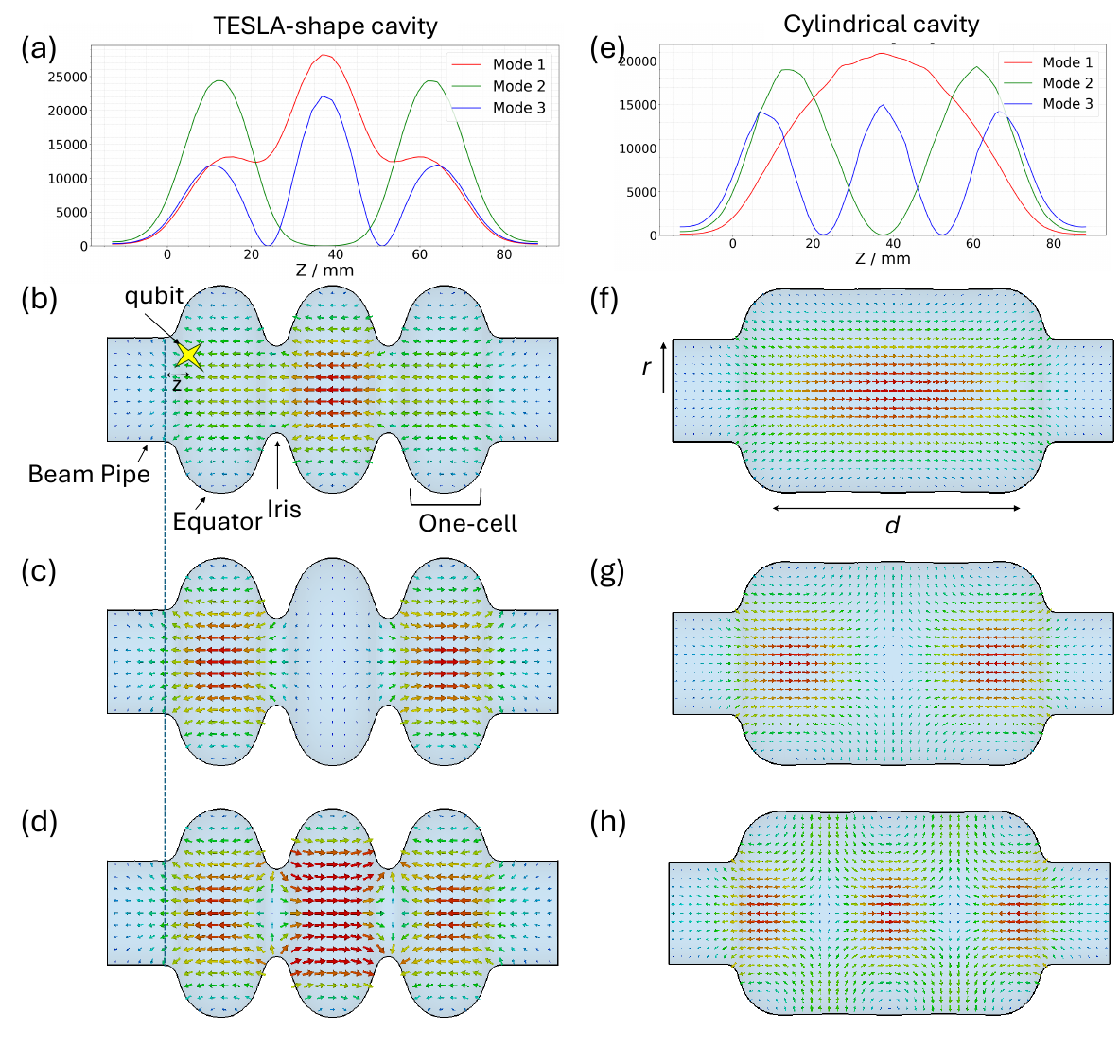}\caption{\justifying Cavity design and finite-element eigenmode simulations of TESLA-shape (a--d) and cylindrical (e--h) SRF cavities. \textbf{Top row (a, e)}: Absolute electric-field magnitude $\left\|\mathrm{E}\right\|$ along the cavity's central axis ($z$, in mm) for the three TM fundamental modes-- Mode 1 (green), Mode 2 (red), and Mode 3 (blue). Each curve shows three peaks corresponding to the three cells. \textbf{Lower three rows (b–d for TESLA, f–h for cylindrical)}: Longitudinal cross-sections of each cavity with the in-plane electric field rendered as a color map (warm colors = high $\left\|\mathrm{E}\right\|$, cool = low), the vector arrows show the field direction. \textbf{(b, f):} The TESLA half-cell geometry (b) is parametrized by the iris radius, equator radius, cell half-length, and two ellipse semi-axes (full construction in App.~\ref{app:tesla-geometry}); the cylindrical cavity (f) is defined by its radius and length. The yellow star shows the position of a transmon within the cavity. The dotted line indicates the cavity opening (cavity entrance), which is the $z=0$ reference shared throughout this work (see Sec.~\ref{sec:method-transmon-dnn}).}\label{fig:figure_cavity}
\end{figure}

In a simple \emph{cylindrical} cavity, there are two classes of modes: the transverse-magnetic (TM) modes, exhibiting a nonzero $E_z$ field and a zero $H_z$ field, and transverse-electric (TE) modes, exhibiting a nonzero $H_z$ field and a zero $E_z$ field~\cite{pozar2011microwave}. In specific configurations, the fundamental $\mathrm{TM}_{010}$ mode shows the highest quality factor and is the ideal candidate for cavity-based quantum computing; its electric field lines run parallel to the cavity's central axis~(Fig.~\ref{fig:figure_cavity}(e--h)). 
In multi-cell cavities (Fig.~\ref{fig:figure_cavity}(a--d)), routinely employed in particle accelerators~\cite{padamsee2023superconducting}, the inter-cell coupling splits the per-cell fundamental mode into as many $\mathrm{TM}_{010}$-like modes as there are cells, forming a passband~\cite{aune2000superconducting}. The mode spacing within the passband is controlled primarily by inter-cell EM coupling, which in turn is dominated by the iris radius (Fig.~\ref{fig:figure_cavity}(b)) of the passage between cells\footnote{Throughout this work, ``iris'' refers to the geometric opening in the cavity wall between adjacent cells (or between a cell and the beam pipe). The reference point used for transmon position in Sec.~\ref{sec:method-transmon-dnn} is the \emph{cavity opening} specifically --- the iris between the beam pipe and the first (or only) cavity cell.}. The smaller the iris radius, the narrower we expect the passband to be. For a three-cell cavity, for example, the $\mathrm{TM}_{010}$-like passband consists of modes differing between each cell by a complex phase of $\pi/3$, $2\pi/3$, and $\pi$, producing respectively 0, 1, and 2 nodes of the real component of $E_z$ axially~\cite{ReineriOptim2023}. A cavity shape of particular relevance here is the \emph{TESLA}\footnote{TESLA stands for \emph{TeV-energy Superconducting Linear Accelerator}: said cavities were originally designed for the TESLA project at DESY research center, Germany.} shape --- an elliptical-cell evolution of the simple cylindrical cavity --- which we use in single-cell form as the fixed reference geometry for the transmon inverse-design dataset (Sec.~\ref{sec:method-transmon-dnn}) and in three-cell form for the EPR characterization (Sec.~\ref{sec:results-epr}); its half-cell construction and accelerator-physics design rationale are summarized in App.~\ref{app:tesla-geometry}. Figure~\ref{fig:figure_cavity} shows a comparison of a three-cell, TESLA-shape cavity (left column), and a cylindrical cavity (right column). The figure also shows the electric-field structure of the lowest-frequency TM modes.

In this work, cylindrical and multicell cavities are parametrized using cubic splines while optimizing for electromagnetic parameters in Sec~\ref{sec:method-cavity-dnn}. Meanwhile, for the transmon inverse-design dataset of Sec.~\ref{sec:method-transmon-dnn}, a single-cell TESLA cavity serves as a fixed, well-characterized reference geometry, decoupled from the spline-parametrized cavities studied in the rest of the work.

Throughout this manuscript, the axial coordinate $z$ is measured from the \emph{cavity opening} - where the beam pipe meets the first (or only) cavity cell - and increases along the cavity axis into the cavity body. The radial coordinate $r$ is the cylindrical distance from the cavity axis. For 3D-cQED applications, the TESLA-shape cavities used here are modified to include beam pipes that accommodate transmon insertion; because these beam pipes can be of various lengths in different physical realizations, we standardize on the cavity opening as $z = 0$, giving a common reference point for both the cavity-DNN spline control points (Sec.~\ref{sec:method-cavity-dnn}) and the transmon-DNN insertion depth $z_\mathrm{trans}$ (Sec.~\ref{sec:method-transmon-dnn}) that is independent of beam-pipe length. The dotted line in Fig.~\ref{fig:figure_cavity}(b) and in Fig.~\ref{fig:control-points} marks this entrance.

\subsection{Cavity QED}
A transmon chip is inserted into the cavity through an aperture, commonly referred to as the beam pipe, and is typically patterned on a silicon or sapphire substrate~\cite{PaikCavity2011}. Figure~\ref{fig:figure_cavity}(b) shows the location of a transmon qubit within the cavity. The cavity--transmon system operates in the dispersive regime~\cite{BlaisReview2021}, in which the qubit--cavity coupling is small compared to the characteristic mode detunings, allowing the system to be described by an effective diagonal Hamiltonian. In second-quantized form, with mode creation and annihilation operators $\{\hat{a}^\dagger_m\}_{m \geq 0}$ and $\{\hat{a}_m\}_{m \geq 0}$ and with the convention $m=0$ for the transmon-like mode, the Hamiltonian reads~\cite{MinevEPR2021}:
\begin{equation}
\begin{aligned}
\hat{H} = \sum_{m \geq 0} \hbar(\omega_m - \Delta_m)\hat{a}_m^{\dagger}\hat{a}_m - \frac{1}{2}\sum_{m \geq 0} \hbar K_m (\hat{a}_m^{\dagger})^2(\hat{a}_m)^2 - \\
\sum_{m > n \geq 0} \hbar \chi_{m,n}\hat{a}_m^{\dagger}\hat{a}_m \hat{a}_n^{\dagger}\hat{a}_n.
\end{aligned}
\label{eq:dispersive-H}
\end{equation}

Here, $\omega_m$ are the linear mode frequencies obtained from the linearized eigenmode solution, $\chi_{m,n}$ are the cross-Kerr nonlinearities, $K_m = \chi_{m,m}/2$ are the self-Kerr coefficients --- the transmon's, $\alpha \equiv K_0 = \chi_{0,0}/2$, being its anharmonicity --- and $\Delta_m = \frac{1}{2}\sum_n \chi_{m,n}$ is the first-order Lamb shift, so that the dressed mode frequency is $\omega_m - \Delta_m$. These nonlinear terms govern dispersive readout, state encoding, mode selectivity, and unwanted crosstalk~\cite{BlaisReview2021}, and are the interactions a 3D-cQED design must control.

To extract the effective Hamiltonian parameters of Eq.~\eqref{eq:dispersive-H} with the use of a classical electromagnetic solver, we employ the EPR method~\cite{MinevEPR2021}. For each mode $m$, the energy participation ratio $p_{mj}$ quantifies the fraction of inductive energy stored in Josephson junction $j$, and these participations control how strongly each mode inherits the junction nonlinearity. Expanding the Josephson potential to fourth order and treating the nonlinear contribution perturbatively yields the Kerr matrix to leading order in the Josephson nonlinearity, with entries
\begin{equation}
\chi_{m,n} \approx \sum_j \frac{\hbar\, \omega_m \omega_n}{4 E_j}\, p_{mj}\, p_{nj},
\label{eq:chi-EPR}
\end{equation}
where the sum runs over all Josephson junctions $j$ and $E_j$ is the Josephson energy of junction $j$. The diagonal entries of this matrix are the self-Kerr coefficients $K_m$ defined above (the transmon's being its anharmonicity $\alpha$), and the off-diagonal entries are the cross-Kerr couplings; higher-order corrections in the Josephson nonlinearity are neglected. The sign convention follows Ref.~\cite{MinevEPR2021}; cross-Kerr magnitudes $|\chi_{m,n}|$ are reported throughout, and the sign of each term in Eq.~\eqref{eq:dispersive-H} is fixed by this convention. In practice, these nonlinear parameters are extracted from the finite-element eigenfields (Ansys HFSS) by \texttt{pyEPR}, which numerically diagonalizes the full nonlinear Josephson Hamiltonian constructed in the EPR framework rather than evaluating the leading-order form of Eq.~\eqref{eq:chi-EPR} directly; these diagonalized values are the source of truth for the coupling parameters reported here and used as the DNN training labels.

Three parameters of practical interest for transmon-mediated control of the cavity are the effective coupling rate $g$ between the transmon and the $\mathrm{TM}_{010}$ cavity mode, the dressed transmon frequency $\nu_q$ (the $g \!\to\! e$ transition frequency, dispersive shifts included; frequencies $\nu$ and angular frequencies $\omega$ are related by $\omega = 2\pi\nu$ throughout), and the transmon anharmonicity $\alpha$ (the self-Kerr $K_0$ above, taken as a magnitude $\alpha = |{-}E_C/\hbar| > 0$, with the sign carried in Eq.~\eqref{eq:dispersive-H}). We adopt $(g, \nu_q, \alpha)$ as the inverse-design targets. They are neither the only parameters of the coupled system nor strictly independent of one another, but they form a compact, weakly redundant set that fixes the operating point. The qubit--cavity cross-Kerr $\chi_{0m}$, in particular, adds little once $(g, \nu_q, \alpha)$ and the cavity frequency $\nu_c$ are set: to leading order in the dispersive expansion it follows from the standard transmon dispersive-shift relation~\cite{KochTransmon2007}, written here in the convention consistent with the \texttt{pyEPR}-extracted cross-Kerr of Eq.~\eqref{eq:chi-EPR} (with $\chi_{0m}>0$),
\begin{equation}
\chi_{0m} \approx \frac{2\,g^2\,\alpha}{\delta\,(\delta - \alpha)},
\label{eq:chi-dispersive}
\end{equation}
where $\delta \equiv \nu_q - \nu_c$ is the qubit--cavity detuning. This same relation makes $\chi_{0m}$ a poor regression target: through the $1/[\delta(\delta-\alpha)]$ factor it varies over orders of magnitude across the design space and grows without bound as the qubit approaches resonance with the cavity ($\delta \to 0$), where the dispersive description itself breaks down, whereas $g$ varies smoothly with geometry. We therefore design against $(g, \nu_q, \alpha)$ and report $\chi_{0m}$ as a derived, physically interpretive quantity.

In the EPR pipeline the logic runs the other way: the numerical diagonalization returns the dressed frequencies, the anharmonicity $\alpha$, and the cross-Kerr $\chi_{0m}$ directly, and the coupling $g$ is recovered by inverting Eq.~\eqref{eq:chi-dispersive},
\begin{equation}
g \approx \sqrt{\left|\frac{\chi_{0m}\,\delta\,(\delta - \alpha)}{2\,\alpha}\right|}.
\label{eq:g-eff}
\end{equation}
Thus $\chi_{0m}$ is what the simulation yields, while $(g, \nu_q, \alpha)$ is the set we design against; Eq.~\eqref{eq:chi-dispersive} is the bridge between the two, used both to turn a desired $\chi_{0m}$ into a target $g$ when a design is specified in terms of the cross-Kerr and to cross-check the extracted couplings (the bad-solve filter of Sec.~\ref{sec:method-transmon-dnn}). Both Eq.~\eqref{eq:chi-dispersive} and Eq.~\eqref{eq:g-eff} hold only in the dispersive regime $g \ll |\delta|$ (the same condition under which $\chi_{0m}$ stays finite), justifying our analysis restriction to $g/|\delta| < 10\%$.

The targets $(g, \nu_q, \alpha)$ are mapped to three transmon \emph{design variables} $(z_\mathrm{trans}, L_J, w_\mathrm{pad})$ --- the junction insertion depth $z_\mathrm{trans}$, the Josephson inductance $L_J$, and the pad width $w_\mathrm{pad}$ --- evaluated from multiphysics simulation, with other choices (substrate material, pad spacing, beam-pipe length) held fixed. These three design parameters were chosen because together they span the operating range of interest and the network learns the mapping accurately (Sec.~\ref{sec:results-transmon-dnn}); their formal definitions, the rationale for each (including why pad \emph{width} rather than length is the varied dimension), and their sampling ranges are given in Sec.~\ref{sec:method-transmon-dnn}, where they are the design variables the DNN predicts.

\section{Methods}
\label{sec:method}

\begin{figure*}[!ht]
\centering
\includegraphics[width=0.9\textwidth]{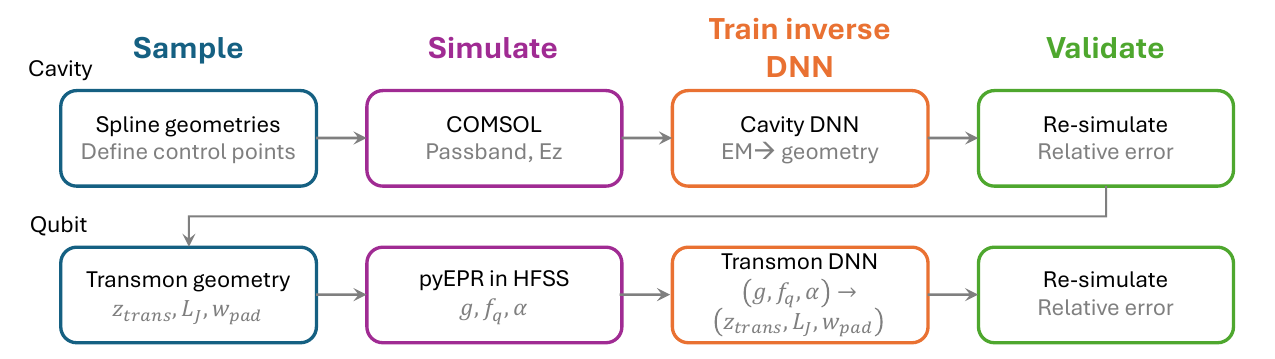}
\caption{\justifying Overview of the two inverse-design pipelines. Both follow the same four-stage, data-driven template: sample candidate geometries, evaluate them with finite-element simulation (the forward map), train a DNN to invert that map, and validate the predicted geometry by re-simulation. The cavity track (top) samples spline-defined cavity profiles, extracts the electromagnetic profile and eigenmodes in COMSOL, and trains a DNN mapping these EM observables to cavity geometry. The transmon track (bottom) samples transmon designs and extracts the qubit--cavity parameters $(g, \nu_q, \alpha)$ via HFSS eigenmode simulation and pyEPR, and trains a DNN mapping those target parameters back to a transmon design.}
\label{fig:pipelines}
\end{figure*}

\begin{figure*}[!ht]
  \centering
  \includegraphics[width=0.7\textwidth]{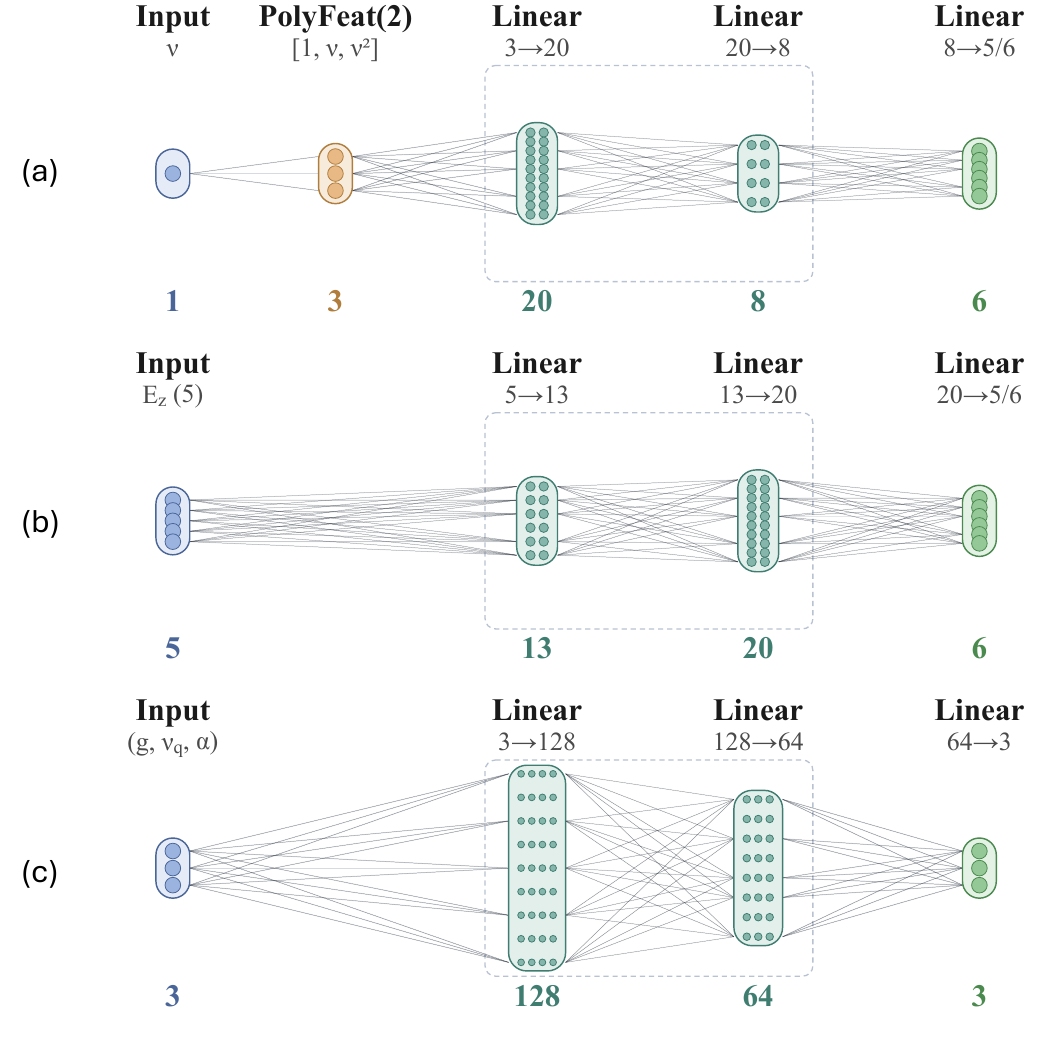}
  \caption{\justifying Architectures of the three feed-forward DNNs used in this work.
    \textbf{(a)}~$f_{\mathrm{freq}}$: scalar $\mathrm{TM}_{010}$ frequency input, expanded by a degree-2 polynomial map $[1,\nu,\nu^{2}]$ and passed through $\mathrm{Linear}(3\!\to\!20\!\to\!8\!\to\!5/6)$, producing the 5 or 6 cavity degrees of freedom for \texttt{cylindrical} or \texttt{multicell} geometries, respectively (Sec.~\ref{sec:method-cavity-dnn}).
    \textbf{(b)}~$f_{\mathrm{field}}$: five-point axial $E_z$ sample at the prospective qubit location, passed through $\mathrm{Linear}(5\!\to\!13\!\to\!20\!\to\!5/6)$.
    \textbf{(c)}~$f_\theta$: transmon inverse-design network mapping target qubit--cavity parameters $(g,\nu_q,\alpha)$ to transmon design variables $(z_\mathrm{trans},L_J,w_\mathrm{pad})$ through $\mathrm{Linear}(3\!\to\!128\!\to\!64\!\to\!3)$ (Sec.~\ref{sec:method-transmon-dnn}).
    Every linear layer is followed by a ReLU activation; hidden layers are grouped by the dashed box.
    Each stadium represents one layer: its height, width, and dot-column count all scale with the layer width, so the over-parametrized hidden layers are visually distinct from the low-dimensional inputs and outputs. Exact widths are printed below each layer; rendered neuron counts are schematic.}
  \label{fig:dnn_architectures}
\end{figure*}

The one-to-many inverse mapping from EM parameters to viable geometric features is well-suited for a data-driven approach. Many perturbations around an initial geometry are randomly sampled, then Maxwell's equations are solved for each one. The EM analysis is made via commercial software packages, such as COMSOL Multiphysics. This way, a labeled dataset of \emph{geometry, EM response} pairs is acquired. A DNN is trained on this dataset to invert the forward map (geometry $\to$ response), taking the response parameters as input and proposing geometries as output. When the targeted observables are not algebraically constrained by one another --- as for the cavity's $E_z$ field and $\mathrm{TM}_{010}$ frequency, or the transmon's $(g, \nu_q, \alpha)$ --- there is no closed-form inverse, and the DNN supplies a learned one; a quantity that is so constrained, such as the cross-Kerr $\chi_{0m}$, is redundant as a target and excluded from this analysis (Sec.~\ref{sec:background}). Because the feedforward DNNs used in this work are deterministic by construction, a given input always produces a single geometry. The choice of training distribution thus determines which solution branch the network learns when the inverse is multi-valued. In the present work, the low validation errors reported in Sec.~\ref{sec:results} suggest that within the sampled design space, each target is matched by essentially one viable geometry. If multiple equivalent geometries existed for a given target, the network would have to compromise between them and the validation error would be substantially larger. Table~\ref{tab:DNN_types} summarizes the DNN input-output pairs used in the work, while the following subsections lay out the individual DNN architectures (illustrated in Fig.~\ref{fig:dnn_architectures}) in detail. Figure~\ref{fig:pipelines} shows the pipelines for the inverse design of both cavity geometry and transmon coupling parameters. Although the two tracks could in principle be coupled (a cavity-DNN geometry feeding the transmon design), in the present work they are run independently: the transmon DNN uses a fixed reference cavity, not a cavity-DNN output.

\begin{table*}[t]
    \centering
    \small
    \caption{Summary of the three DNN types used in the work. The first two tackle cavity geometry inverse design, while the third tackles transmon inverse design.}
    \begin{tabular}{@{}p{0.16\textwidth}p{0.23\textwidth}p{0.23\textwidth}p{0.27\textwidth}@{}}
    \toprule
    DNN type & Frequency-inputting\newline cavity design & $E_z$ field-inputting \newline cavity design & Transmon design \\
    \midrule
    \midrule
    Input & $\mathrm{TM}_{010}$-related \newline frequency & $E_z$ field at transmon location & Transmon parameters\newline $(g, \nu_q, \alpha)$ \\
    \midrule
    Output & Cavity wall geometry & Cavity wall geometry & Transmon geometry\newline $(z_\mathrm{trans}, L_J, w_\mathrm{pad})$\\
    \bottomrule
    \end{tabular}
    \label{tab:DNN_types}
\end{table*}

\subsection{Geometry parametrization and data collection}
\label{sec:method-geometry}

Cavity geometries must be parametrized in a form suitable for DNN training. The standard elliptical-cell parametrization in the SRF literature represents each cell's meridional contour by two mutually tangential ellipses (see App.~\ref{app:tesla-geometry})~\cite{PadamseeBook2009,Aune2000TESLA}. A less common alternative defines cells using splines \cite{RiemannSpline2012}. Splines can generate more general shapes at the cost of direct control over some geometrical parameters, such as the iris radius or cell width. Still, using splines is preferable: accessing a wider variety of shapes is a decisive factor for the DNN training. Sampling proceeds from a generic initial curve, defined by specifying the coordinates of all control points. The initial geometry is set heuristically so its eigenmodes fall into the microwave regime. The degrees of freedom, i.e.\ the coordinates of the splines' control points, are sampled from a uniform distribution around the values of the initial curve. The width of the uniform distribution varies depending on the control point.

\begin{figure*}[htbp]
\centering
\includegraphics[width=0.9\textwidth]{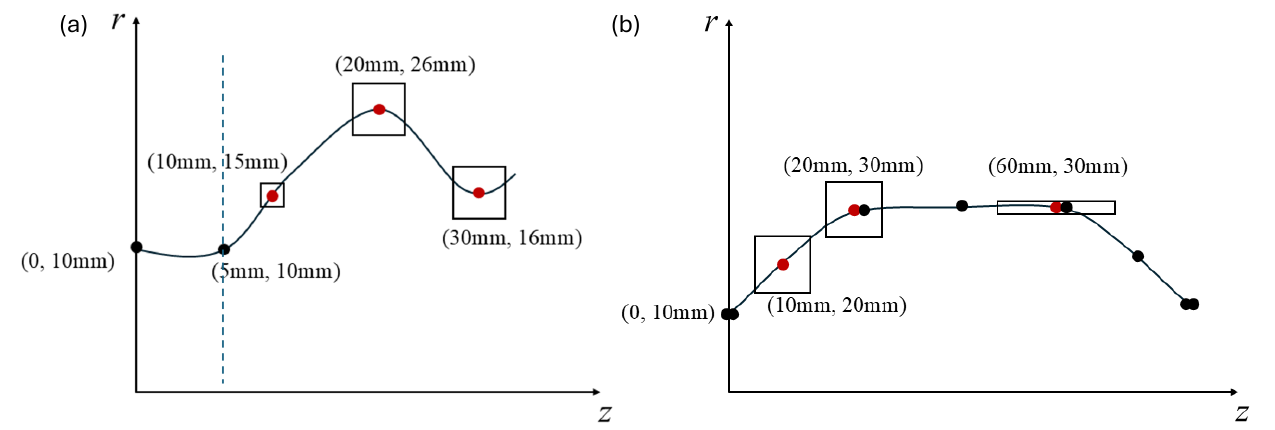}

\caption{\justifying (a) Curve generation for \texttt{multicell} geometries. Coordinates give the locations of the control points of the initial curve. Rectangles represent uniform sampling regions. For the second and third red control points the distribution width is 8 mm, while for the first it's 1.8 mm due to its proximity to the cavity opening. This structure is repeated for the specified number of cells. The dotted vertical line in panel (a) marks the cavity opening (cavity entrance), which serves as the $z = 0$ reference shared with Fig.~\ref{fig:figure_cavity} and with the transmon-side insertion depth $z_\mathrm{trans}$ (Sec.~\ref{sec:method-transmon-dnn}). (b) Curve generation for \texttt{cylindrical} geometries. The first two red control points are sampled with widths of 8 mm. The narrow rectangle corresponds to horizontal sampling of the length only, with width 16 mm, done after the positions of the last two degrees of freedom are fixed. In both geometry types, the sampling widths are chosen heuristically to allow for maximal robustness while avoiding self-intersections, pointed edges, and fundamental mode frequencies outside the microwave regime.}
\label{fig:control-points}
\end{figure*}

The curves are further separated into two geometry types. The first, which we call \texttt{multicell}, uses a repeating control-point structure and mimics the construction of a multi-cell, TESLA-like cavity. Within one cell, there are six degrees of freedom: the axial and radial coordinates of select control points. The second geometry type has no repeating structure. We call this type \texttt{cylindrical}. Though less general, this generation method allows for more careful control of geometrical characteristics. In this type of generation, there are five degrees of freedom: the axial and radial coordinates of two select control points, as well as the cylinder length (difference in the axial coordinates of the two following control points). Both geometry generation methods are detailed in Fig.~\ref{fig:control-points}, with sampled control points, corresponding to the degrees of freedom, highlighted in red, while the fixed ones are black. Finally, cylindrical ``beam pipes'' are appended to both ends of the cavity. The diameter is such to ensure the electric field is evanescent at the boundary, mirroring physical realizations.

For this work, the used \texttt{multicell} initial curve is 3-celled with a fundamental frequency of $5.43~$GHz, while the \texttt{cylindrical} initial curve fundamental frequency is $4.09~$GHz. Results reporting on the mode analysis are in Sec. \ref{sec:results}. The boundary conditions in COMSOL are set to perfect conductors, with the cavity interior modeled as vacuum. For the cavity inverse-design problem of Sec. \ref{sec:method-cavity-dnn}, a dataset of 2000 3-celled \texttt{multicell} geometries and 2000 \texttt{cylindrical} geometries is generated and run through COMSOL to extract the relevant EM parameters. Each data point required 2 minutes of COMSOL computational time.

For the transmon inverse-design dataset of Sec.~\ref{sec:method-transmon-dnn}, the chosen cavity is a single-cell TESLA-shape geometry with its $\mathrm{TM}_{010}$ mode at $\nu_c \approx 5.06$~GHz, providing a fixed, well-isolated mode dispersively coupled to the transmon. 
The choice of target $(g, \nu_q, \alpha)$ --- and the resulting cross-Kerr $\chi_{0m}$ between transmon and cavity, fixed by the target via Eq.~\eqref{eq:chi-dispersive} --- is motivated by the intended operation. Small $\chi_{0m}$ in the sub-MHz range is favored for storing quantum information in long-lived cavity modes with minimal inverse Purcell decay through the qubit, as exploited in recent multi-qudit cavity-based experiments~\cite{KimSQMS2506}. Conversely, larger $\chi_{0m}$ approaching $\sim 1$~MHz enables faster ancilla-mediated control such as the SNAP gate~\cite{HeeresSNAP2015}, at the cost of stronger qubit--cavity hybridization. The transmon inverse-design DNN of Sec.~\ref{sec:method-transmon-dnn} targets the dispersive-regime $(g, \nu_q, \alpha)$ widely used in bosonic control with superconducting qubits, with the corresponding $\chi_{0m}$ spanning both the storage and fast-control regimes above.

\subsection{Cavity geometry inverse-design DNN: dataset and architecture}
\label{sec:method-cavity-dnn}

For the 3-celled \texttt{multicell} geometry type, DNNs were trained to take as input one of the following vectors of EM parameters:
\begin{itemize}
\item The difference in frequency between the $\pi$ and $\pi/3$ $\mathrm{TM}_{010}$-like modes that defines the passband width. We target the passband width because widening it resolves the frequency crowding and transmon-induced crosstalk of the three $\mathrm{TM}_{010}$ modes~\cite{ReineriSQMS2308}. The mean frequency of the three passband modes is not constrained as an input; it varies across the dataset, bounded to the microwave range by the spline-sampling widths of Sec.~\ref{sec:method-geometry}. Imposing a target mean frequency would require adding it as a second input, outside the scope of this work.
\item The averaged real part of the normalized $E_z$ field across the $\pi/3$, $2\pi/3$, and $\pi$ $\mathrm{TM}_{010}$-like modes, evaluated at fixed axial sample points along the cavity.
We set control points at -5~mm, -2~mm, 1~mm, 4~mm, and 7~mm axially and 5~mm radially from the cavity opening (here, the axial coordinate $z = 0$ baseline corresponds to the dotted line as in Fig. \ref{fig:control-points}, while the radial coordinate corresponds exactly to $r$ in Fig. \ref{fig:control-points}). These fixed coordinates were chosen to align with the transmon location inside the cavity. The $E_z$ field is the input here because its $z$-component magnitude directly dictates the cavity-transmon electric dipole coupling. Using $E_z$ from the bare cavity (i.e., with the transmon dielectric and metallization absent from the COMSOL eigenmode solve) as the DNN target is justified by the dispersive operating regime of Sec.~\ref{sec:background}: the transmon's back-action on the cavity field is perturbatively small, so the bare $E_z$ at the prospective qubit location closely approximates the dressed field that determines the coupling. The three modes share the same transverse field character --- $E_z$ parallel to the cavity axis --- and differ only in their inter-cell phase advance ($\pi/3$, $2\pi/3$, $\pi$), which sets the number of axial sign-flips of $E_z$ along the cavity (0, 1, and 2 respectively)~\cite{ReineriOptim2023}; the averaged input therefore captures the common transverse envelope while suppressing the phase-dependent axial variation between modes.

\end{itemize}

Similarly, for \texttt{cylindrical} geometries, DNNs were trained on inputs of one of the following vectors of EM parameters:
\begin{itemize}
\item The frequency of the fundamental $\mathrm{TM}_{010}$ mode, which influences the transmon-cavity coupling through Eq.~(\ref{eq:chi-dispersive}).
\item The real part of the normalized $E_z$ field of the $\mathrm{TM}_{010}$ mode at 0, 3~mm, 6~mm, 9~mm, and 12~mm axially and 5~mm radially from the cavity opening (here, the axial and radial coordinates correspond exactly to $z$ and $r$ in Fig. \ref{fig:control-points}). The same reasoning as in the \texttt{multicell} case applies here and at these fixed coordinates. This time no averaging of the $E_z$ field is done, as cylindrical cavities have a unique $\mathrm{TM}_{010}$ mode.
\end{itemize}

ReLU\footnote{ReLU stands for Rectified Linear Unit, and is defined as: $\mathrm{ReLU}(x) = 0$ for $x < 0$ and $\mathrm{ReLU}(x) = x$ for $x \geq 0$.} was used as the activation function in all hidden layers \cite{Glorot2011ReLU}.
The DNN architecture (number of hidden layers and units per layer) was selected by a small grid search under the heuristic constraint that the training-set size exceeds the total parameter count by a factor of roughly $3$--$4$, with hidden layers wider than the input/output layers for stable regression performance. The wide hidden layers, consisting of a linear transform followed by a ReLU activation function, serve to express the desired properties of our target mappings. In each layer, the linear transform takes in a vector of nonnegative features. This is because the input to each hidden layer is the output of the previous one, naturally nonnegative due to the ReLU function properties. The linear transform takes the vector of nonnegative features into a vector of both positive and negative features, while the hidden layer's ReLU activation function sets the negative features to zero. In this way, a hidden unit expresses a property of the final desired mapping onto its inputs. Combining many wide ReLU hidden layers yields a universal function approximator~\cite{Hornik_1989}. For the DNNs whose first layer inputs the $E_z$ field at 5 points around the transmon location, for both geometry types, the architecture used was:
\begin{equation}
    f_{\mathrm{field}}: (5 \ E_z \ \mathrm{points}) \longmapsto (5 \ \mathrm{or} \ 6 \ \mathrm{cavity \ geometry \ DoF}),
\end{equation}

\begin{equation}
\begin{aligned}
    \mathrm{[Linear(5 \rightarrow 13),\ ReLU,\ Linear(13 \rightarrow 20),\ } \\
    \mathrm{ReLU,\ Linear(20 \rightarrow 5 \ or \ 6)]}.
\end{aligned}
\end{equation}
For the DNNs whose first layer inputs the $\mathrm{TM_{010}}$ passband width or frequency, depending on the geometry type, the exact architecture used was

\begin{equation}
\begin{aligned}
    f_{\mathrm{freq}}: (\mathrm{TM_{010} \ bandwidth \ or \ frequency}) \\
    \longmapsto (\mathrm{5 \ or \ 6 \ cavity \ geometry \ DoF})
\end{aligned}
\end{equation}

\begin{equation}
\begin{aligned}
    \mathrm{[PolynomialFeatures(2),\ Linear(3 \rightarrow 20),\ ReLU,} \\ \mathrm{\ Linear(20 \rightarrow 8),\
    ReLU,\ Linear(8 \rightarrow 5 \ or \ 6)]}.
\end{aligned}
\end{equation}

Here, the notation $\mathrm{Linear}(a\rightarrow b)$ means the linear layer transforms $a$ input nodes into $b$ output nodes, while $\mathrm{PolynomialFeatures}(n)$ denotes an $n$-dimensional polynomial map\footnote{In our case, where the input vector is one-dimensional and corresponds to a frequency value $\nu$, the quadratic polynomial map yields the vector $\left[1,\ \nu,\ \nu^2\right]^T$}. In the architectures, 6 features are output in the final layer for \texttt{multicell} geometries (one per cell degree of freedom — the structure then repeats across the three cells), while 5 features are output for \texttt{cylindrical} geometries, corresponding to the number of degrees of freedom in their control points. A second-degree polynomial map was applied to the initial input in the frequency-type architectures, allowing the network to capture a potential quadratic dependence. Finally, mean square error (MSE) was used as the loss metric for DNN backpropagation and performance assessment:
\begin{equation}
    \mathrm{MSE} = \frac{1}{N}\sum_{j = 1}^N (\vec{x_j}^{\mathrm{actual}} - \vec{x_j}^{\mathrm{guess}})^2.
\end{equation}
Here, $N$ is the number of data points considered, while $\vec{x_j}^{\mathrm{actual}}$ and $\vec{x_j}^{\mathrm{guess}}$ are the vectors of degrees of freedom of individual data points. In the \texttt{multicell} case, these vectors are six-dimensional, while in the \texttt{cylindrical} case they are five-dimensional.

The datasets of 2000 points for both geometry types were partitioned by a random $80/20$ split into a training and testing set. To assess the performance of the DNNs, three quality checks were performed: 
\begin{enumerate}
    \item verify that training and testing losses decrease and converge;
    \item compare geometries from the testing set with predicted DNN output geometries;
    \item run predicted geometries back through COMSOL to compare the resulting EM parameters against the original DNN inputs corresponding to data points from the testing set.
\end{enumerate}
These checks were applied to all DNN configurations described above.

\subsection{EPR characterization of DNN-generated cavity geometries}
\label{sec:method-epr}
This subsection details a workflow to characterize the coupling properties of cavity DNN outputs via the EPR method. To this end, the \texttt{multicell} passband-width DNN is queried with two target passband widths. The resulting cavity geometries are exported to HFSS, a simplified transmon is inserted, and EPR analysis is performed via \texttt{pyEPR} across a range of transmon insertion depths. This yields the dispersive Hamiltonian parameters [Eq.~\eqref{eq:chi-EPR}] for each cavity as a function of insertion depth, characterizing the cavity DNN outputs in terms of their transmon-control parameters. This characterization workflow is independent of the transmon inverse-design DNN of Sec.~\ref{sec:method-transmon-dnn}, which holds its cavity fixed rather than drawing it from the cavity DNN.

\subsection{Transmon inverse-design DNN: dataset and architecture}
\label{sec:method-transmon-dnn}

The transmon inverse-design DNN of this subsection addresses a second, independent inverse-design problem at the transmon level. In contrast to the cavity-geometry problem of Sec.~\ref{sec:method-cavity-dnn}, which is itself tackled by more than one network (Table~\ref{tab:DNN_types}), here the cavity is held fixed and the network maps target qubit--cavity parameters $(g, \nu_q, \alpha)$ to the three transmon design variables $(z_\mathrm{trans}, L_J, w_\mathrm{pad})$ introduced in Sec.~\ref{sec:background}. We solve this inverse problem with a dedicated DNN trained on HFSS eigenmode simulations processed through the \texttt{pyEPR} framework.

The three design variables output by the DNN, illustrated in Fig.~\ref{fig:multicell_transmon_cad}, are defined formally as:
\begin{itemize}
\item $z_\mathrm{trans}$: axial position of the Josephson junction along the cavity axis, measured from the \emph{cavity opening} (i.e., the $z = 0$ reference of Sec.~\ref{sec:background} marked by the dotted line in Fig.~\ref{fig:figure_cavity}(b)). Positive values place the junction inside the cavity body; small or near-zero values keep it within the beam-pipe region, where it couples to the evanescent field outside the cavity proper. This primarily tunes the field overlap between the junction and the cavity modes and is the leading determinant of the coupling rate $g$;
\item $L_J$: Josephson inductance, which sets the junction energy $E_J = (\Phi_0/2\pi)^2 / L_J$ and, jointly with the pad capacitance, determines the transmon frequency $\nu_q$ and anharmonicity $\alpha$ through the $E_J/E_C$ ratio~\cite{KochTransmon2007};
\item $w_\mathrm{pad}$: width of the two superconducting pads forming the transmon shunt capacitance, which sets the charging energy $E_C$ via $C_\Sigma$ and therefore co-determines $\nu_q$ and $\alpha$ alongside $L_J$. As discussed in Sec.~\ref{sec:background}, width (rather than length) was chosen as the variable pad dimension so that the junction center remains at a fixed axial position relative to the chip. Pad geometry also affects $g$.
\end{itemize}

The three design variables $(z_\mathrm{trans}, L_J, w_\mathrm{pad})$ defined above were drawn independently from uniform distributions over the full ranges $z_\mathrm{trans} \in [0.71,\, 13.71]$~mm, $L_J \in [7,\, 20]$~nH, and $w_\mathrm{pad} \in [0.25,\, 2.0]$~mm, with later batches adding manually stratified sub-regions targeting identified coverage gaps in parameter space. These ranges span the operating regimes of interest for 3D-cQED: $z_\mathrm{trans}$ runs from junctions near the cavity opening (weak coupling) to deeper insertion, where the junction samples a larger $\mathrm{TM}_{010}$ field amplitude; $L_J$ covers inductances that, combined with the pad capacitance, keep $E_J/E_C$ in the transmon regime; $w_\mathrm{pad}$ spans pad widths producing typical 3D-cQED qubit frequencies of order a few GHz.
All simulations were performed with the transmon inserted into the single-cell TESLA-shape cavity ($\mathrm{TM}_{010}$ at $\approx 5.06$~GHz), which presents a cleaner, well-separated $\mathrm{TM}_{010}$ mode than the three-cell cavities of the EPR characterization (Sec.~\ref{sec:results-epr}).

The dataset was assembled in successive rounds of simulation, with later rounds targeting sparse regions of parameter space, such as high $\alpha$, high $|\chi|$, and the sparsely-sampled region where $\alpha < 75$~MHz and $g > 35$~MHz, identified from earlier rounds. Two independent estimates of $g$ are extracted from each HFSS+\texttt{pyEPR} solution. The first, $g_\mathrm{PR}$, is computed directly from the junction energy-participation ratios via the participation-based vacuum coupling of Ref.~\cite{MinevEPR2021}, without diagonalizing the nonlinear Hamiltonian. The second, $g$, is obtained from the \emph{diagonalized} cross-Kerr [Eq.~\eqref{eq:chi-EPR}] by inverting the dispersive relation [Eq.~\eqref{eq:chi-dispersive}, i.e.\ via Eq.~\eqref{eq:g-eff}]. Because the two draw on different parts of the solve, their agreement within $10\%$ indicates a clean solve. Post-hoc filters were applied to the simulated dataset to remove bad solves, using three physics-based criteria:
\begin{enumerate}
    \item $|g - g_\mathrm{PR}|/g > 10\%$, indicating mode mixing or an avoided crossing;
    \item $g > 200$~MHz, indicating spurious mode identification;
    \item $g/|\delta| > 10\%$, indicating breakdown of the dispersive approximation.
\end{enumerate}
After applying this filter, $981$ clean points remain out of $1100$ total simulated.

As established in Sec.~\ref{sec:background}, the cross-Kerr $\chi_{0m}$ is excluded as a network input; consistent with that reasoning, supplying it as an additional input did not improve accuracy in practice. The architecture is a fully connected feedforward network:

\begin{equation}
f_\theta : (g, \nu_q, \alpha) \longmapsto (z_\mathrm{trans}, L_J, w_\mathrm{pad}),
\label{eq:transmon-dnn-symbolic}
\end{equation}
\begin{equation}
\begin{aligned}
[\mathrm{Linear}(3 \to 128),\ \mathrm{ReLU},\ \mathrm{Linear}(128 \to 64),\ \\
\mathrm{ReLU},\ \mathrm{Linear}(64 \to 3)],
\end{aligned}
\end{equation}
with hidden widths $128$ and $64$ chosen heuristically to keep the parameter count one to two orders of magnitude above the training-set size, and with hidden layers wider than the three-feature input and output for stable regression performance. All six input/output features were independently standardized to zero mean and unit variance using training-set statistics only. Training used Adam~\cite{kingma2015adam} with learning rate $10^{-3}$ and weight decay\footnote{Both are dimensionless coefficients in the Adam update rule: the learning rate scales each gradient step in the standardized parameter space, and the weight decay is the L2-regularization multiplier.} $10^{-4}$, batch size $32$, and up to $2000$ epochs; these settings were selected heuristically and held fixed across runs. The model state at the epoch of lowest held-out test loss was retained as the final checkpoint.

Following the validation protocol used for the cavity DNN of Sec.~\ref{sec:method-cavity-dnn}, generalization was assessed by an $80/20$ train/test split ($784$ training points, $197$ held out), repeated over $15$ independent random shuffles of the $981$-point clean dataset; the shuffle yielding the lowest held-out test MSE was retained as the representative training run, and its per-dimension relative root mean squared error (RRMSE) values are reported in Sec.~\ref{sec:results-transmon-dnn}.
The primary performance metric is RRMSE per output dimension:
\begin{equation}
\mathrm{RRMSE}_k = \frac{\mathrm{RMSE}_k}{y_k^{\max} - y_k^{\min}} \times 100\%,
\end{equation}
where $y_k^{\max} - y_k^{\min}$ is the full output range over the training set. In addition to this held-out evaluation, the DNN-predicted designs were re-simulated in HFSS with EPR extraction to verify end-to-end recovery of the target parameters, mirroring the COMSOL round-trip check applied to the cavity DNN.

\section{Results}
\label{sec:results}

The results that follow are split into three distinct problems that together span the full cavity-qubit design stack. Section~\ref{sec:results-cavity-dnn} reports the cavity inverse-design DNN, which maps target electromagnetic observables (passband width and Ez profile) to a cavity geometry. The design is validated by COMSOL re-simulations. Section~\ref{sec:results-epr} addresses the forward map of two cavity geometries produced by the cavity DNN. EPR is used to extract how the cross-Kerr couplings vary with the insertion depth of the transmon qubit in the resonant cavity. Section~\ref{sec:results-transmon-dnn} returns to the inverse design, now at the transmon level. With the cavity geometry fixed, we map target qubit-cavity parameters $(g, \nu_q, \alpha)$ into transmon geometries $(z_{trans}, L_J, w_{\mathrm{pad}})$. These results are validated by HFSS simulations.

\subsection{Cavity DNN performance}
\label{sec:results-cavity-dnn}

\begin{figure*}[!htbp]
\centering
\includegraphics[width=\linewidth]{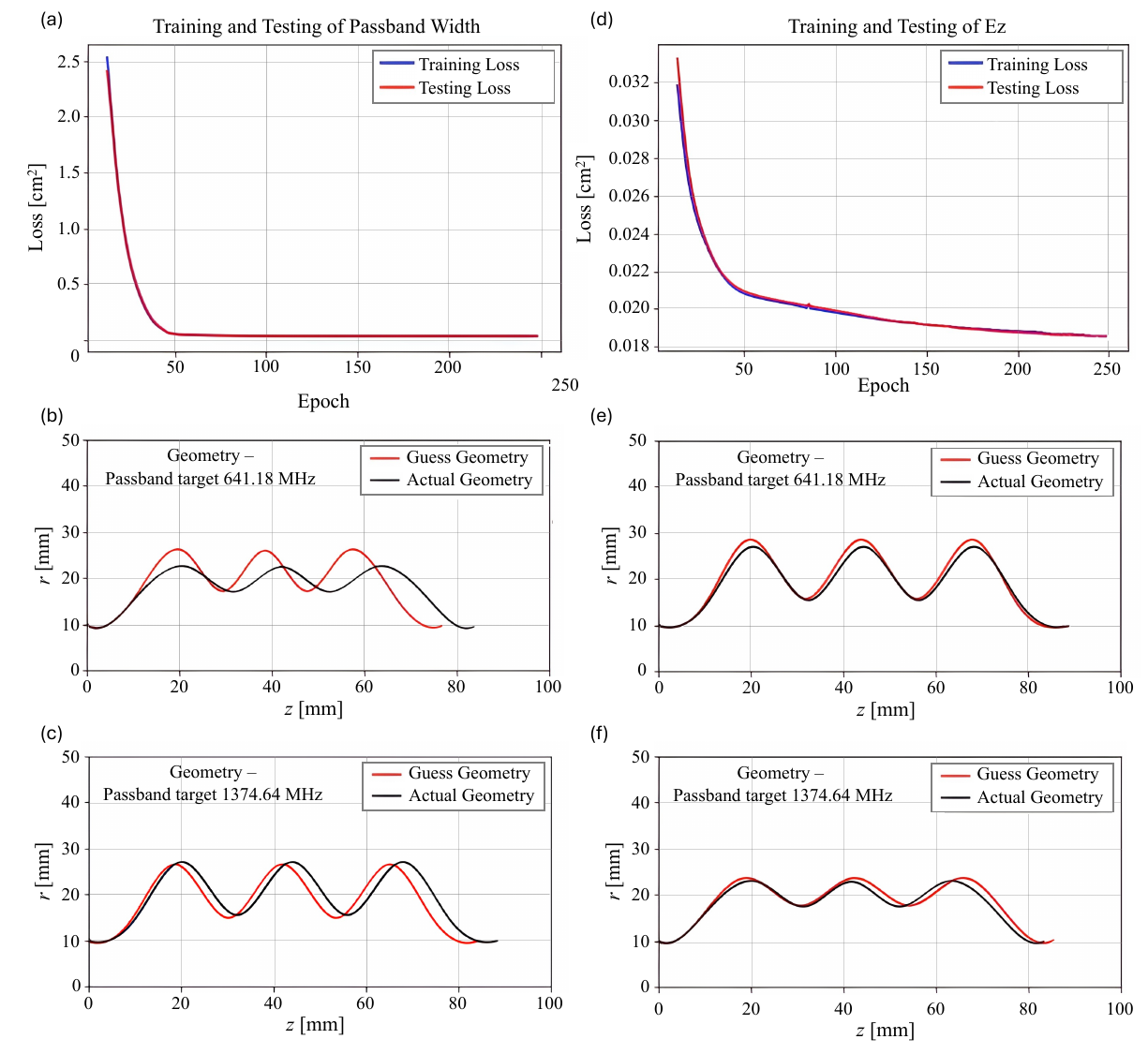}
\caption{\justifying Training and testing results for the \texttt{multicell} geometry. \textbf{(a)} Training and testing MSE loss of the passband width model from epoch~15 onward. The trend for previous epochs shows a sharp drop from around 300~cm$^2$; \textbf{(b-c)} Example output (guess) geometry of the passband width model vs.\ actual. (b) Widths are 659.33~MHz (actual) and 641.18~MHz (guess). (c) Second example output (guess) geometry vs.\ actual. Widths are 1374.64~MHz (actual) and 1380.88~MHz (guess). \textbf{(d)} Training and testing loss of the 5-point $E_z$ model from epoch 15 onward. The trend for previous epochs shows a sharp drop from around 4~cm$^2$. \textbf{(e-f)} Example output (guess) geometries of the 5-point $E_z$ model vs.\ actual for both examples. }
\label{fig:training_testing}
\end{figure*}

Before constructing the cavity DNNs, we verified that the initial curves used to seed the cavity-geometry sampling behave as expected. 
The \texttt{multicell} initial curve exhibits the expected $\mathrm{TM}_{010}$-like passband of three modes (0, 1, and 2 axial $E_z$ nodes), lying $> 3$~GHz below the next TE mode. 
Complete COMSOL mode spectra for both reference geometries are reported in App.~\ref{app:pillbox}. 
This section describes a representative subset of results produced by our DNN algorithms for the \texttt{multicell} geometry type (Fig.~\ref{fig:training_testing}). Analogous results for the \texttt{cylindrical} geometry are deferred to App.~\ref{app:pillbox}. 

\subsubsection*{DNN input: $\mathrm{TM}_{010}$-like passband width}

The \texttt{multicell} passband-width DNN converges with the MSE loss falling from approximately $300$~cm$^2$ to about $0.05$~cm$^2$ \footnote{Note that for cavity geometry matching, we chose to present the absolute loss and not the usual percentage drop from the initial loss. This way, it serves also as a comparative metric between the overall geometrical matching capabilities of the frequency DNN and the $E_z$ field DNN.}, as shown in Fig.~
\ref{fig:training_testing}(a) from epoch 15 onward. This represents a drop of about $99.98 \%$.

The DNN's output geometries deviate from the held-out reference geometries in regions that weakly affect the passband width, indicating that the network has learned to focus on the geometric features most relevant to the target output. The actual (input) versus predicted passband width values match closely, with relative errors of at most $\sim5 \%$ for the considered examples. This is consistent with the network prioritizing geometric features that dominate the output. Specifically, the iris radii of the predicted vs.\ actual geometries match to within $\sim5 \%$ as well for the considered examples, in accordance with the passband width being highly dependent on inter-cell coupling. A visual representation of guessed and actual geometries with minor passband variations due to similar dominant geometric features is reported in Fig.~
\ref{fig:training_testing}(b-c).

\subsubsection*{DNN input: 5 $E_z$ points around the transmon}

The \texttt{multicell} 5-point field model converges with training and testing MSE losses dropping from approximately $4$~cm$^2$ to $0.02$~cm$^2$, as shown in Fig.~
\ref{fig:training_testing}(d) from epoch 15 onward. The lower converged absolute loss value suggests better geometrical matching overall. This represents a relative drop of about $99.5\%$.

\begin{figure*}[!htbp]
\centering
\includegraphics[width=\linewidth]{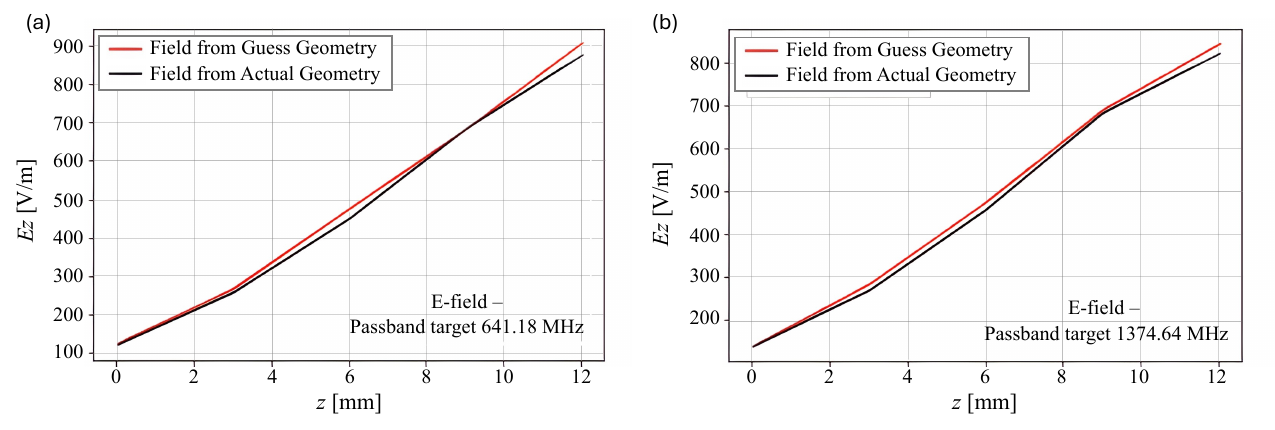}
\caption{\justifying (a) Example (guess) field produced by the DNN output vs.\ DNN input field. (b) Second example (guess) field produced by the DNN output vs.\ DNN input field.}
\label{fig:multicell-field-fields}
\end{figure*}

With this model, geometry matching improves relative to the passband-width model, as the lower converged loss value would suggest.
The DNN output varies in both overall size and detailed shape, in contrast to the previous case where the iris radius dominated. When the predicted geometries are run back through COMSOL, each point of the resulting $E_z$ distributions matches the input field's corresponding point to at most about $\sim 4\%$ for the considered examples. Fig.~
\ref{fig:training_testing}(e-f) shows the same two example geometries from the previous subsection, while Fig.~\ref{fig:multicell-field-fields} shows the fields they produce.

\subsection{EPR results}
\label{sec:results-epr}

To explore how the $\mathrm{TM}_{010}$-like passband width affects transmon--cavity coupling parameters, we performed two EPR characterization studies, each starting from a different cavity produced by the cavity-geometry DNN of Sec.~\ref{sec:method-cavity-dnn}. The \texttt{multicell} passband-width DNN was queried with two target passband widths of 400~MHz and 800~MHz. The generated geometries yielded passband widths of 382.15~MHz and 831.95~MHz when re-simulated in COMSOL --- within 5\% of the requested targets in both cases.

\begin{figure}[htbp]
\centering
\includegraphics[height=0.3\textheight,keepaspectratio]{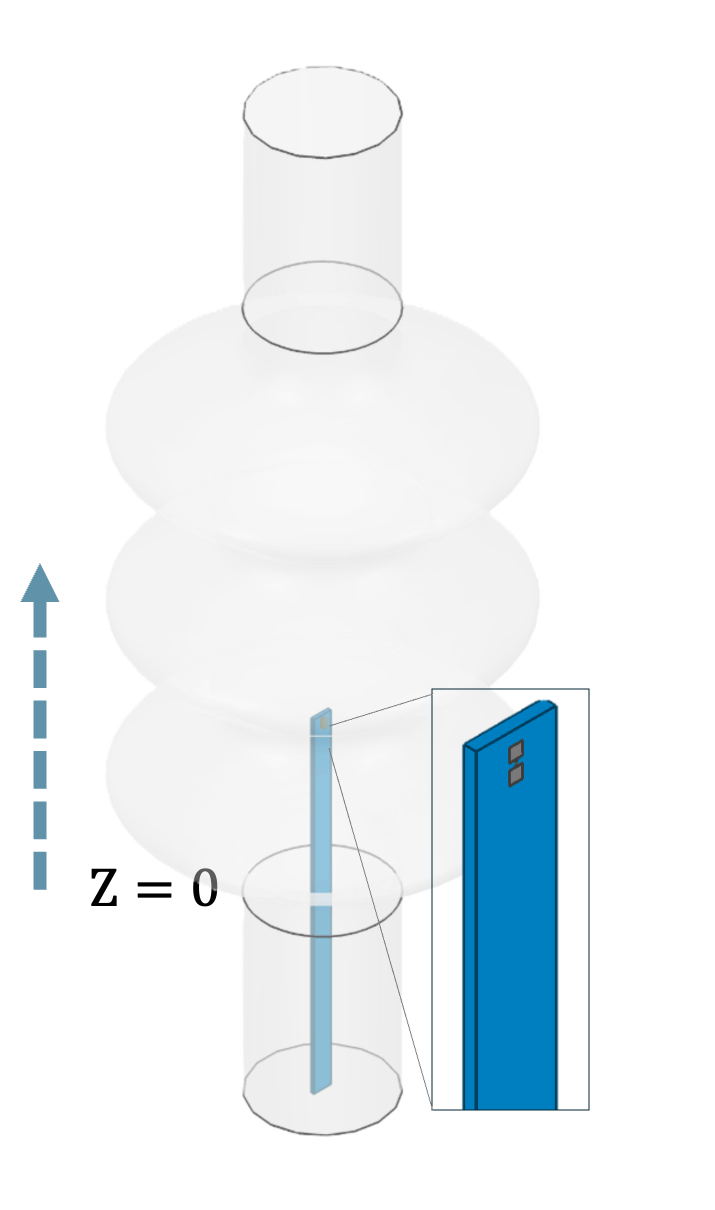}
\caption{\justifying CAD model of the 3-cell multicell cavity with an inserted transmon and a close-up of the chip itself: two superconducting capacitor pads of width $w_\mathrm{pad}$ on the dielectric substrate, connected by the lumped Josephson inductance $L_J$. The Josephson junction sits in the beam pipe at axial position $z_\mathrm{trans}$, measured from the cavity opening. The same transmon-chip geometry was used in both the 400~MHz and 800~MHz cavities --- which differ only in passband width --- for the EPR characterization of Sec.~\ref{sec:results-epr}.}
\label{fig:multicell_transmon_cad}
\end{figure}

Both generated cavities were exported to HFSS, and the transmon CAD model was inserted along the cavity axis at radial offset $r = 5$~mm. The junction's position relative to the cavity opening was swept over $z_\mathrm{trans} \in [13.88, 21.38]$~mm for the 400~MHz cavity and $z_\mathrm{trans} \in [15.78, 23.28]$~mm for the 800~MHz cavity, using the convention defined in Sec.~\ref{sec:method-transmon-dnn}. Larger $z_\mathrm{trans}$ corresponds to the junction sitting deeper inside the cavity body, and to a longer substrate on which the circuit elements are patterned (Fig.~\ref{fig:multicell_transmon_cad}). For each insertion depth, an HFSS eigenmode simulation was performed, and \texttt{pyEPR} was used to extract the participation ratios $p_{mj}$ for the three $\mathrm{TM}_{010}$-like passband modes. The corresponding nonlinear parameters were then obtained from these participations via \texttt{pyEPR}, following the EPR method of Sec.~\ref{sec:background} [Eq.~\eqref{eq:chi-EPR}]. Appendix~\ref{app:cad} describes the transmon--cavity CAD model common to both the single-cell and multicell (Fig.~\ref{fig:multicell_transmon_cad}) cavities used in this work.

\begin{figure*}[htbp]
\centering
\includegraphics[width=0.8\textwidth]{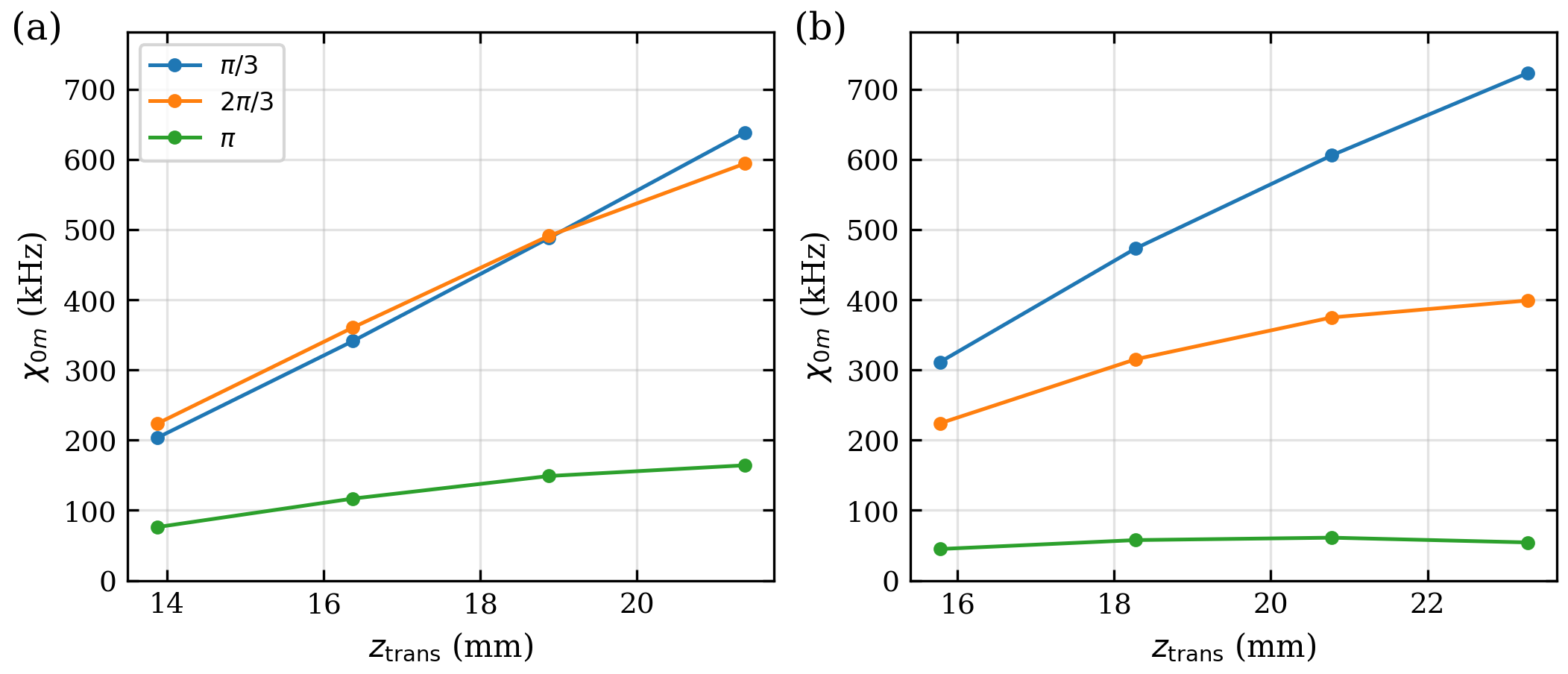}
\caption{\justifying (a) Cross-Kerr coefficients $\chi_{0m}$ for the $\pi/3$, $2\pi/3$, and $\pi$ modes of the 400~MHz cavity. (b) Corresponding coefficients for the 800~MHz cavity.}
\label{fig:cross-kerr}
\end{figure*}

The extracted nonlinear parameters behave as expected: dispersive coupling increases as the transmon is inserted farther into the cavity, since the $\mathrm{TM}_{010}$-like passband modes acquire larger participation in the junction. Figure~\ref{fig:cross-kerr} shows the transmon--mode cross-Kerr coefficients for the $\pi/3$, $2\pi/3$, and $\pi$ modes of the passband for both the 400~MHz and 800~MHz geometries. In both cases, the couplings increase monotonically with insertion depth. In the broader-passband geometry, the larger iris and stronger inter-cell coupling delocalize the three passband modes across the cells, so that $E_z$ at the fixed transmon location differs more sharply between these modes, producing the quantitative differences in cross-Kerr seen between the two panels of Fig.~\ref{fig:cross-kerr}. 

\subsection{Transmon inverse-design DNN performance}
\label{sec:results-transmon-dnn}

Whereas Sec.~\ref{sec:results-epr} \emph{characterized} fixed cavities by reading off their EPR cross-Kerr as a function of insertion depth --- the natural forward observable (Fig.~\ref{fig:cross-kerr}) --- we now address the \emph{inverse} problem of mapping a target operating point to a transmon design. Following Sec.~\ref{sec:background}, we specify the target by $(g, \nu_q, \alpha)$ rather than by the cross-Kerr (Eq. (\ref{eq:chi-dispersive})).

\subsubsection*{Train/test performance}
\label{sec:cv-perf}

\begin{figure}[htbp]
\centering
\includegraphics[width=\myfigwidth]{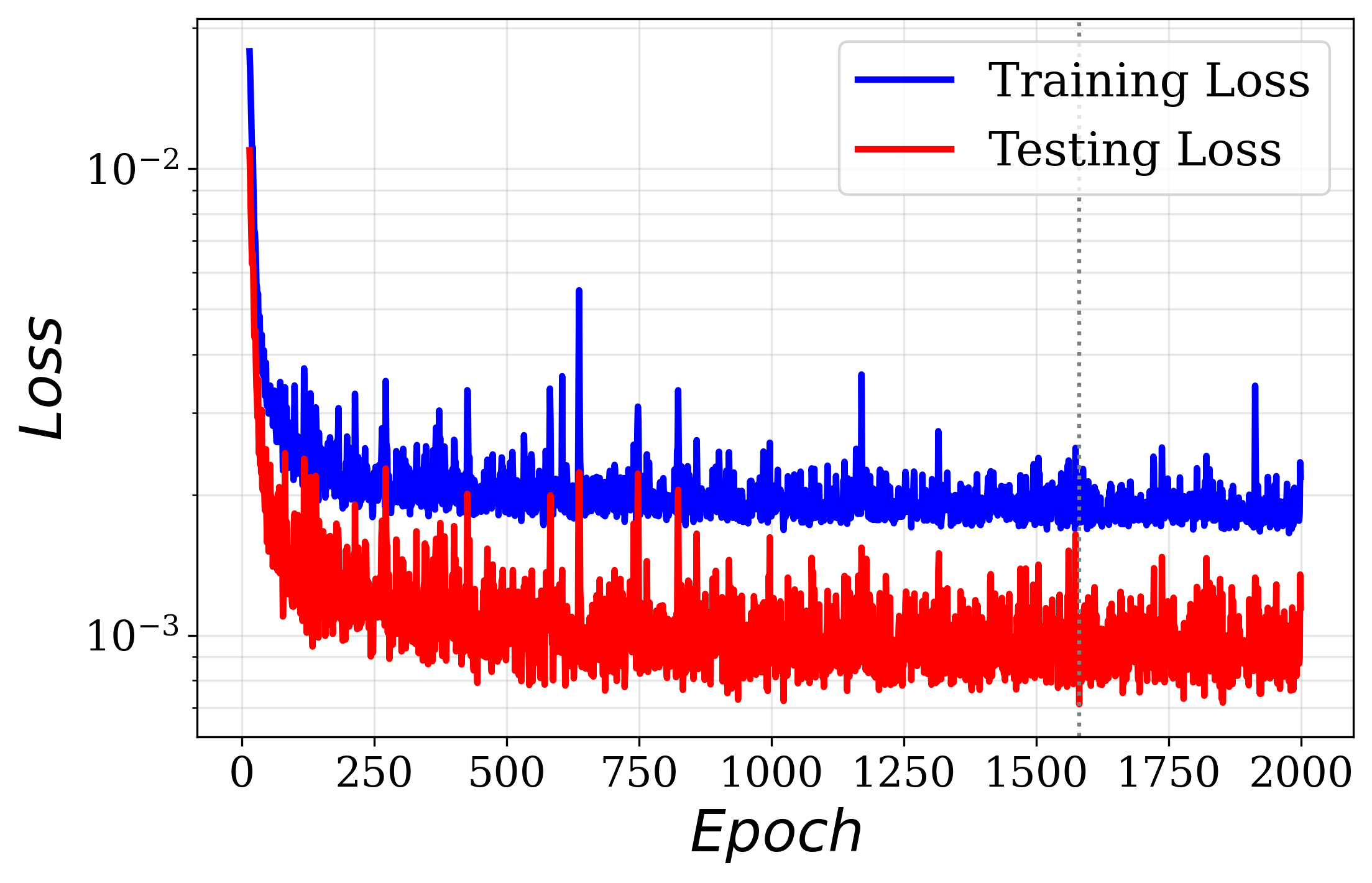}
\caption{\justifying Training and testing loss curves of the transmon inverse-design DNN under the $80/20$ split, shown from epoch~$15$ onward to make the post-initialization drop visible. The vertical dotted line marks the epoch of lowest test loss, at which the model checkpoint is taken.}
\label{fig:cv-loss}
\end{figure}

Figure~\ref{fig:cv-loss} shows the training and testing loss curves of the transmon inverse-design DNN. The curves converge and track each other without overfitting, and the model state at the epoch of lowest test loss is retained as the final checkpoint (marked by the dotted vertical line).

Across the $15$ shuffles the per-dimension RRMSE averages $1.81 \pm 0.96\%$. Table~\ref{tab:rrmse} reports the breakdown for the representative run --- the shuffle with the lowest held-out MSE, which reaches a mean RRMSE of $0.72\%$ and is therefore a best case rather than a typical one; because each shuffle's checkpoint is chosen on its own held-out loss, the held-out figures are in any case mildly optimistic. Across the three outputs, this run reproduces the held-out reference designs with absolute RMSE of $\approx 0.09$~mm in $z_\mathrm{trans}$, $\approx 0.06$~nH in $L_J$, and $\approx 0.018$~mm in $w_\mathrm{pad}$; the range-normalized RRMSE values reflect the relative sizes of the three sampling ranges ($\sim 13$~mm, $\sim 13$~nH, and $\sim 1.75$~mm respectively). The $w_\mathrm{pad}$ RRMSE of $1.02\%$ is the largest despite the smallest absolute residual, an artifact of the narrow sampling range and the range-normalization convention.

\begin{widetab}[t]
\centering
\small
\setlength{\tabcolsep}{7pt}
\caption{Per-dimension held-out RRMSE for the representative (lowest-MSE) shuffle; see text for the across-shuffle statistics.}
\label{tab:rrmse}
\begin{tabular}{lcl}
\toprule
Output & RRMSE (\%)  & Note \\
\midrule
$z_\mathrm{trans}$ & $0.71$ & -
 \\
$L_J$              & $0.43$ & -
 \\
$w_\mathrm{pad}$   & $1.02$ & range-norm artifact \\
\midrule
\textbf{Mean}      & $0.72$ 
& unweighted mean over 3 outputs \\
\bottomrule
\end{tabular}
\end{widetab}

\subsubsection*{Representative test-set examples}
\label{sec:representative-examples}

Two representative held-out test points illustrate the per-point behavior of the DNN as an end-to-end round trip. Table~\ref{tab:examples} lists each target operating point (\emph{DNN input}), the design the DNN proposes for it (\emph{DNN-predicted}), and the parameters recovered when the corresponding geometry is re-simulated in HFSS via EPR (\emph{back-run}). The two points were drawn from the held-out test set: Point 1 lies near the center of the training distribution in standardized $(g, \nu_q, \alpha)$ space and Point 2 sits near a sparser, high-$\alpha$ edge. The cross-Kerr $\chi_{0m}$ implied by each target through the dispersive relation [Eq.~\eqref{eq:chi-dispersive}] is listed alongside it to indicate the operating regime, though $\chi_{0m}$ is not itself a network coordinate: Point~1, at $\chi_{0m} \approx 1$~MHz, sits near the fast-control (SNAP) regime~\cite{HeeresSNAP2015}, while Point~2, an order of magnitude lower at $\approx 0.1$~MHz, lies toward the long-storage regime~\cite{KimSQMS2506}; the two points thus span an order of magnitude in cross-Kerr.

\begin{widetab}[t]
\centering
\footnotesize
\setlength{\tabcolsep}{7pt}
\caption{End-to-end round trip for two representative held-out test points. Each target operating point (\emph{DNN input}) is mapped by the network to a design (\emph{DNN-predicted}), which is re-simulated in HFSS with EPR extraction (\emph{back-run}); parenthesized values are the signed relative deviation of the back-run coupling from its target. The cross-Kerr $\chi_{0m}$ is derived from each target $(g, \nu_q, \alpha)$ via the dispersive relation [Eq.~\eqref{eq:chi-dispersive}] and listed as operating-regime context, not as a network coordinate. Point 1 lies in a dense region of the training distribution; Point 2 sits near a sparse, high-$\alpha$ edge. Design-variable accuracy over the full test set is reported in Table~\ref{tab:rrmse} and App.~\ref{app:parity}.}
\label{tab:examples}
\begin{tabular}{lcc}
\toprule
& Point 1 & Point 2 \\
\midrule
\multicolumn{3}{l}{\emph{Target (DNN input)}} \\
\quad $g$ (MHz)                  & $19.51$    & $20.99$ \\
\quad $\nu_q$ (GHz)              & $4.782$    & $6.677$ \\
\quad $\alpha$ (MHz)             & $169.15$   & $282.08$ \\
\quad $\chi_{0m}$ (MHz, derived) & $1.04$     & $0.115$  \\
\midrule
\multicolumn{3}{l}{\emph{DNN-predicted design}} \\
\quad $z_\mathrm{trans}$ (mm) & $7.910$   & $12.148$ \\
\quad $L_J$ (nH)              & $8.912$   & $7.441$  \\
\quad $w_\mathrm{pad}$ (mm)   & $0.6786$  & $0.2694$ \\
\midrule
\multicolumn{3}{l}{\emph{Back-run EPR of predicted geometry}} \\
\quad $g$ (MHz)              & $19.33~(-0.9\%)$   & $21.07~(+0.4\%)$  \\
\quad $\nu_q$ (GHz)          & $4.805~(+0.5\%)$   & $6.707~(+0.4\%)$  \\
\quad $\alpha$ (MHz)         & $171.83~(+1.6\%)$  & $284.37~(+0.8\%)$ \\
\bottomrule
\end{tabular}
\end{widetab}

Across the three design variables, $L_J$ shows the tightest parity scatter (lowest range-normalized error), while $z_\mathrm{trans}$ tracks closely, consistent with its $r \approx 0.90$ correlation with $g$ (App.~\ref{app:parity}); per-dimension errors are given in Table~\ref{tab:rrmse}. Re-simulating the predicted designs in HFSS with \texttt{pyEPR} closes the round trip: the recovered $(g, \nu_q, \alpha)$ deviate from their targets by at most $\sim 1.6\%$ (on $\alpha$), with $g$ within $\sim 1\%$ and $\nu_q$ within $0.5\%$, as tabulated in the back-run rows of Table~\ref{tab:examples}. Both the typical Point~1 and the edge Point~2 recover all three coordinates to within $\sim 2\%$, confirming that the predicted designs reproduce their targeted Hamiltonian parameters end-to-end across the dense and sparse regions of the training distribution.

\section{Conclusions}
\label{sec:conclusions}

We have presented a deep-learning approach to the inverse design of 3D cQED hardware that operates at two levels of the design stack: (i) SRF cavity geometry and (ii) transmon qubit position and properties. On the cavity side, the DNN architectures generated geometries matching the target passband widths within $\sim 5\%$,
diverging from the held-out reference geometry only in features that weakly affect passband width. This is evidence that the network learned the iris architecture and coupling rather than the full geometry. The 5-point $E_z$ DNN matched both the predicted geometry and the resulting $E_z$ distribution, point by point, within $\sim 4\%$. The transmon inverse-design DNN mapped target qubit--cavity parameters $(g, \nu_q, \alpha)$ to transmon design variables $(z_\mathrm{trans}, L_J, w_\mathrm{pad})$ with a mean held-out RRMSE of $1.81 \pm 0.96\%$ across 15 random shuffles (best-shuffle $0.72\%$) under the $80/20$ train/test protocol. HFSS re-simulation of the DNN-predicted designs recovered the target $(g, \nu_q, \alpha)$ to within $\sim 2\%$ for representative test-set points spanning two distinct operating regimes, confirming that the predicted designs reproduce their targeted Hamiltonian parameters end-to-end.
These models provide a fast design path to candidate geometries, replacing repeated manual parameter sweeps that each take several minutes and accumulate across the many iterations needed for a full design. Because the transmon network is trained on coupling parameters extracted by the EPR method via \texttt{pyEPR}~\cite{MinevEPR2021}, the approach embeds a machine-learning inverse model within the standard quantum-device characterization workflow, inverting the geometry-dependent set of self- and cross-Kerr interactions that is impractical to tune by hand. The same strategy should extend to other quantum-device design problems where the geometry-to-Hamiltonian map is analytically intractable and the design space is high-dimensional and one-to-many.
Future work includes extending the input feature set to richer observables, both for the cavity, e.g., higher order modes and loss parameters, and the qubits, e.g., extending coverage to the low-$\alpha$ and high-$g$ regions of parameter space, as well as exploring alternative geometry parametrization. Extending the approach to broader design spaces where multiple viable geometries coexist for a given target would also be valuable; this would require replacing the deterministic feedforward networks used here with architectures capable of producing several candidate geometries per input. More broadly, the deterministic feedforward networks used here could be replaced by generative or invertible architectures that return multiple candidate designs per target (quantum computing applications), addressing qubits weakly or strongly coupled to the cavity, e.g.,  favoring long-lived storage or fast ancilla-mediated controls, respectively.


\backmatter

\section*{Author contributions}
S.Z. and D.M.K. conceived the project and supervised the research. J.M. and J.Y. contributed equally to the work, focusing on the inverse design of the cavity and qubit, respectively. A.R. contributed to the development of the pyEPR code wrapper and the initial setup of the microwave simulations. J.Y and S.Z. contributed to the final editing of the manuscript. All authors reviewed the manuscript and approved the final version. 

\section*{Data availability}
Datasets, Python scripts, and COMSOL simulations used in the work will be made available upon request to the authors.

\section*{Competing Interests} 
The authors declare no competing interests.

\section*{Acknowledgements}
Fermilab is operated by Fermi Forward Discovery Group, LLC under Contract No. 89243024CSC000002 with the U.S. Department of Energy, Office of High Energy Physics. This work was supported by the U.S. Department of Energy, Office of Science, National Quantum Information Science Research Centers, Superconducting Quantum Materials and Systems Center (SQMS), under Contract No. 89243024CSC000002. The authors thank Dr. Yao Lu for reviewing the manuscript, and Dr. Sara Sussman and Olivia Seidel for insightful discussions on the inverse design of superconducting quantum systems.

\clearpage

\appendix \section{TESLA-shape cell geometry}
\label{app:tesla-geometry}

The TESLA shape is an evolution of the simple cylindrical cavity, originally developed for high-gradient SRF accelerator cavities in linear-collider technology with a profile chosen to minimize the peak surface electric and magnetic fields relative to the accelerating gradient, thereby enabling both the high quality factors $Q_0$ and the high acceleration gradients required for the application. In this work, single-cell and three-cell TESLA cavities serve as reference and characterization geometries for the transmon DNN dataset (Sec.~\ref{sec:method-transmon-dnn}) and the EPR coupling study (Sec.~\ref{sec:results-epr}).

\begin{figure}[htbp]
\centering
\includegraphics[width=\myfigwidth]{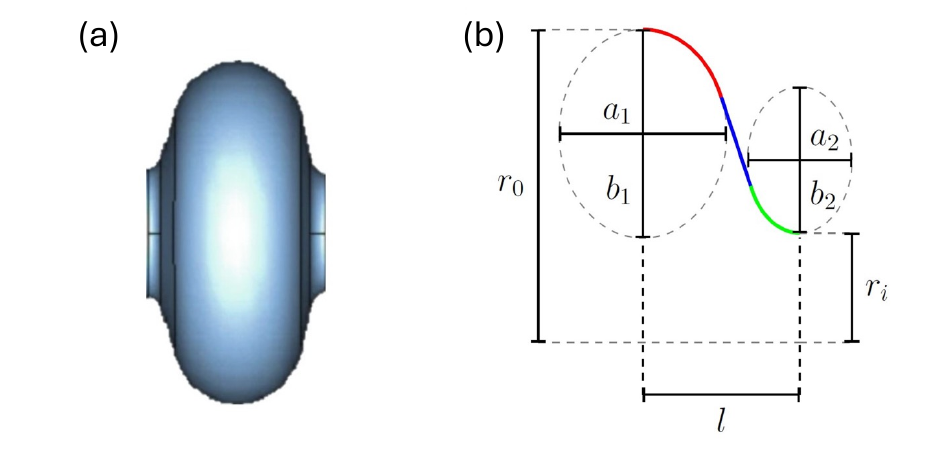}
\caption{\textbf{(a)} TESLA-shaped single-cell cavity CAD. \textbf{(b)} Half-cell outline displaying the two ellipses arcs and the common tangent highlighted in red, green, and blue, respectively.}
\label{fig:single-cell_TESLA}
\end{figure}

Each half-cell is constructed from two mutually tangent ellipses joined by a straight tangent line, parametrized by the iris radius, equator radius, cell half-length, and the semi-axes of the two ellipses (Fig.~\ref{fig:single-cell_TESLA}); full cells are formed by mirroring the half-cell about the equator plane, and $N$ such cells are joined to produce an $N$-cell cavity.

\section{Initial-curve mode spectra and \texttt{cylindrical} DNN results}
\label{app:pillbox}

This appendix first reports the COMSOL eigenmode spectra of the \texttt{multicell} and \texttt{cylindrical} initial curves that anchor the cavity-DNN sampling distributions (Sec.~\ref{sec:method-cavity-dnn}), then presents the DNN performance for the \texttt{cylindrical} geometry.

Table~\ref{tab:multicell-modes} reports the first ten modes of the \texttt{multicell} initial curve in COMSOL. The first three modes --- Modes~1, 2, and 3 --- show 0, 1, and 2 axial nodes of the $E_z$ field, respectively. They are identified as the $\pi/3$, $2\pi/3$, and $\pi$ phase-advance-per-cell members of the $\mathrm{TM}_{010}$-like passband (Sec.~\ref{sec:background}). Their adjacent-mode spacings are $\approx 360$~MHz (Mode~1~$\to$~Mode~2) and $\approx 660$~MHz (Mode~2~$\to$~Mode~3). The next mode, at a higher frequency, is a TE mode ($E_z = 0$).

\begin{widetab}
\centering
\caption{First ten modes of the \texttt{multicell} initial curve in COMSOL.}
\small
\setlength{\tabcolsep}{5pt}
\label{tab:multicell-modes}
\begin{tabular}{cccccc}
\toprule
Mode 1 & Mode 2 & Mode 3 & Mode 4 & Mode 5 & Unit \\
\midrule
5.434 & 5.798 & 6.455 & 9.947 & 10.41 & GHz \\
\midrule
Mode 6 & Mode 7 & Mode 8 & Mode 9 & Mode 10 & Unit \\
\midrule
10.52 & 10.58 & 10.80 & 10.81 & 11.17 & GHz \\
\bottomrule
\end{tabular}
\end{widetab}

Similarly, the \texttt{cylindrical} cavity was initially run through COMSOL's eigenmode solver. The first ten modes are recorded in Table~\ref{tab:pillbox-modes}. The first three modes with nonzero $E_z$ once again exhibit 0, 1, and 2 axial nodes of the $E_z$ field, respectively. Their frequencies are significantly spread compared to the \texttt{multicell}. 
The third lowest mode in frequency is a TE mode lying between $\mathrm{TM}_{011}$ and $\mathrm{TM}_{012}$.

\begin{widetab}
\centering
\caption{First ten modes of the \texttt{cylindrical} initial curve in COMSOL.}
\label{tab:pillbox-modes}
\small
\setlength{\tabcolsep}{5pt}
\begin{tabular}{cccccc}
\toprule
Mode 1 & Mode 2 & Mode 3 & Mode 4 & Mode 5 & Unit \\
\midrule
4.087 & 5.055 & 6.644 & 6.743 & 7.865 & GHz\\
\midrule
Mode 6 & Mode 7 & Mode 8 & Mode 9 & Mode 10 & Unit \\
\midrule
8.673 & 8.982 & 9.479 & 9.637 & 9.928 & GHz \\
\bottomrule
\end{tabular}
\end{widetab}

\begin{figure*}[!htbp]
\centering
\includegraphics[width=\linewidth]{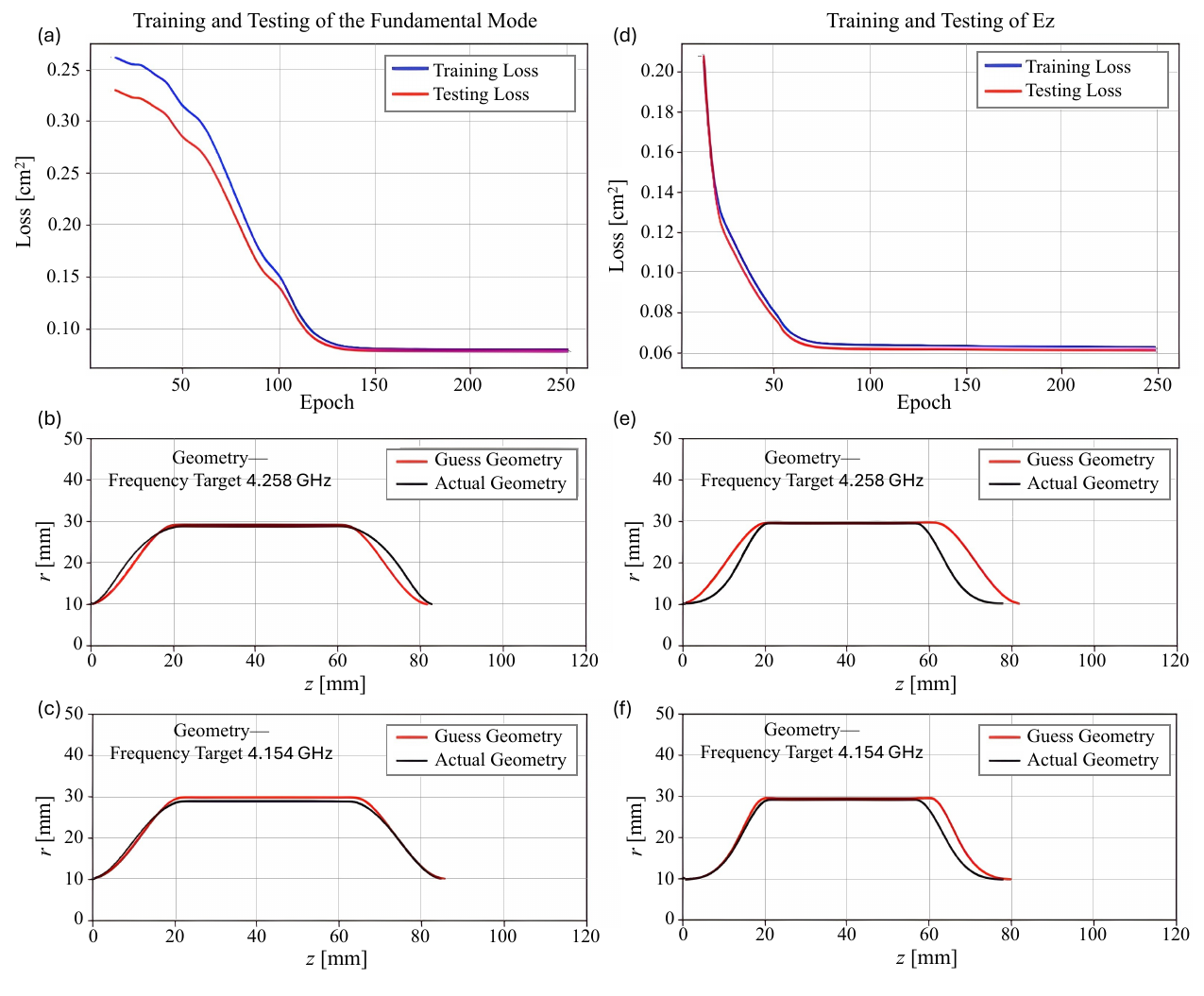}
\caption{\justifying Training and testing results for the \texttt{cylindrical} geometry. \textbf{(a)} Training and testing loss of the fundamental mode model from epoch~15 onward; \textbf{(b-c)} Example output (guess) cylindrical geometry of the fundamental mode model vs. actual. (b) Frequencies 4.265 GHz (actual) and 4.258 GHz (guess). (c) Second example. Frequencies 4.171 GHz (actual) and 4.154 GHz (guess); \textbf{(d)} Training and testing loss of the 5-point $E_z$ model from epoch 15 onward. \textbf{(e-f)} Example output (guess) geometries of the 5-point $E_z$ model vs.\ actual for both examples. }
\label{fig:training_testing_cyl}
\end{figure*}

\subsubsection*{DNN Input: $\mathrm{TM}_{010}$ mode frequency}



The training and testing losses for the fundamental-mode model converge from approximately $3000~\mathrm{cm}^2$ to about $0.08~\mathrm{cm}^2$, as shown in Fig.~\ref{fig:training_testing_cyl}(a) from epoch 15 onward.

The model's performance in matching whole geometries is limited: the network appears to have learned that the fundamental mode frequency depends primarily on the radial width of the cavity. Frequency matching is very precise, with an accuracy within about $\sim 0.5 \%$. Two examples are shown in Fig.~\ref{fig:training_testing_cyl}(b-c).

\subsubsection*{DNN Input: 5 $E_z$ points around the transmon}

The 5-point $E_z$ model converges from approximately $7~\mathrm{cm}^2$ to about $0.06~\mathrm{cm}^2$, as shown in Fig.~\ref{fig:training_testing_cyl}(d) from epoch 15 onward. The lower final converged loss suggests better overall geometry matching for the $E_z$ model.

\begin{figure*}[!htbp]
\centering
\includegraphics[width=\linewidth]{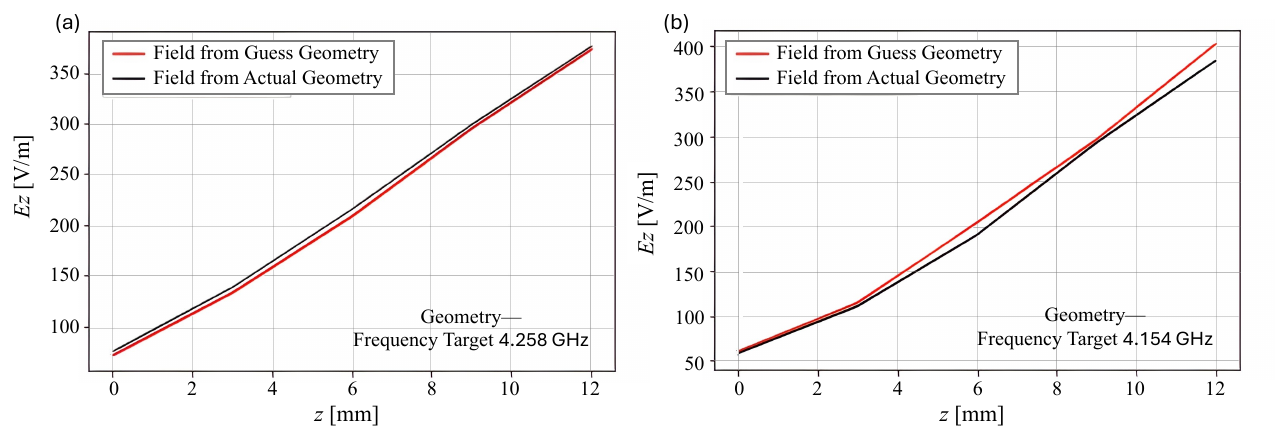}
\caption{(a) Example (guess) field produced by the \texttt{cylindrical} DNN output vs.\ DNN input field. (b) Second example (guess) field vs.\ DNN input field.}
\label{fig:pillbox-field-fields}
\end{figure*}

Two example cavities show better geometrical matching than the frequency model. When back-run through COMSOL, the $E_z$ field from the guess geometry at any considered point does not differ from the one produced by the actual geometry by more than about $\sim 5 \%$. This is shown in Figs.~\ref{fig:training_testing_cyl}(e-f) and~\ref{fig:pillbox-field-fields}(a-b).

\section{CAD model of the transmon--cavity system}
\label{app:cad}

Within the EPR framework, the Josephson junction is modeled as a lumped, purely inductive element ($L_J$) inserted into an otherwise linear electromagnetic structure, shunted by the two capacitor pads of width $w_\mathrm{pad}$ that set the transmon capacitance (Fig.~\ref{fig:multicell_transmon_cad}). For each junction position $z_\mathrm{trans}$, an HFSS eigenmode solve is passed to \texttt{pyEPR} for extraction of the participation ratios and nonlinear Hamiltonian parameters. The same chip geometry is used across both cavity contexts 
--- the single-cell TESLA cavity of the transmon-DNN dataset (Sec.~\ref{sec:method-transmon-dnn}) and the three-cell \texttt{multicell} cavities of the EPR characterization (Sec.~\ref{sec:results-epr}); only the surrounding cavity differs.

\section{Parity plots for the transmon inverse-design DNN}
\label{app:parity}

\begin{figure*}[!htbp]
\centering
\includegraphics[width=0.8\linewidth]{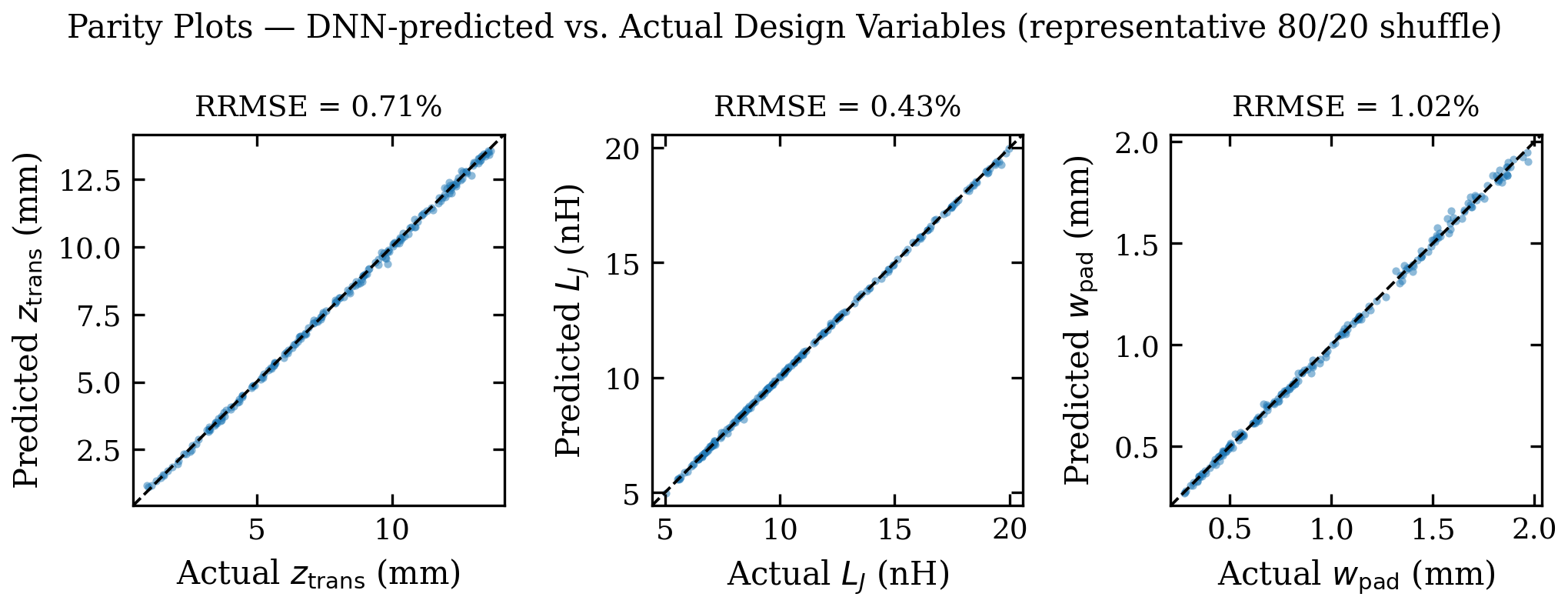}
\caption{Parity plots for DNN-predicted vs.\ actual design variables $(z_\mathrm{trans}, L_J, w_\mathrm{pad})$ for the test set of the representative $80/20$ shuffle. Dashed diagonals indicate perfect prediction. No systematic bias is observed across any dimension.}
\label{fig:parity}
\end{figure*}

Figure~\ref{fig:parity} shows parity plots comparing DNN-predicted and actual design-variable values for all test-set points of the representative $80/20$ shuffle (canonical seed). Parity scatter is tightest for $L_J$, which has the lowest range-normalized error; $z_\mathrm{trans}$ also tracks closely, consistent with its strong linear correlation with $g$ in the training set ($r \approx 0.90$). $w_\mathrm{pad}$ shows the largest range-normalized error.

\clearpage
\bibliographystyle{apsrev4-2}
\bibliography{apssamp}

\end{document}